\documentclass[sigconf]{acmart}

\usepackage[utf8]{inputenc}%(only for the pdftex engine)
\setcopyright{none}
\renewcommand\footnotetextcopyrightpermission[1]{} % removes footnote with conference information in first column
\usepackage{booktabs} % For formal tables
\usepackage{subfig}
\usepackage{colortbl}
\usepackage{multirow}
\usepackage{xcolor}
\usepackage{amsfonts}
\usepackage{amsthm}
\usepackage{graphicx}

\begin{document}

\author{Sameera Horawalavithana}
%\authornote{Dr.~Trovato insisted his name be first.}
% \orcid{1234-5678-9012}
\affiliation{%
 \institution{University of South Florida}
 %\streetaddress{P.O. Box 1212}
 \city{Tampa} 
 \state{FL} 
 %\postcode{43017-6221}
}
\email{sameera1@mail.usf.edu}

\author{Clayton Gandy}
%\authornote{Dr.~Trovato insisted his name be first.}
% \orcid{1234-5678-9012}
\affiliation{%
 \institution{University of South Florida}
 %\streetaddress{P.O. Box 1212}
 \city{Tampa} 
 \state{FL} 
 %\postcode{43017-6221}
}
\email{cgandy@mail.usf.edu}

\author{Juan Arroyo Flores}
%\authornote{Dr.~Trovato insisted his name be first.}
% \orcid{1234-5678-9012}
\affiliation{%
 \institution{University of South Florida}
 %\streetaddress{P.O. Box 1212}
 \city{Tampa} 
 \state{FL} 
 %\postcode{43017-6221}
}
\email{jga@mail.usf.edu}

\author{John Skvoretz}
%\authornote{Dr.~Trovato insisted his name be first.}
% \orcid{1234-5678-9012}
\affiliation{%
 \institution{University of South Florida}
 %\streetaddress{P.O. Box 1212}
 \city{Tampa} 
 \state{FL} 
 %\postcode{43017-6221}
}
\email{jskvoretz@usf.edu}

\author{Adriana Iamnitchi}
%\authornote{The secretary disavows any knowledge of this author's actions.}
\affiliation{%
	\institution{University of South Florida}
	%\streetaddress{P.O. Box 1212}
	\city{Tampa} 
	\state{FL} 
	%\postcode{43017-6221}
}
\email{anda@cse.usf.edu}

\title{Diversity, Topology, and the Risk of Node Re-identification in Labeled Social Graphs}

% \runningtitle{Diversity, Topology, and the Risk of Node Re-identification in Labeled Social Graphs}

  %\subtitle{...}

  \begin{abstract}
{Real network datasets provide significant benefits for understanding phenomena such as information diffusion or network evolution. 
Yet the privacy risks raised from sharing real graph datasets, even when stripped of user identity information, are significant. 
When nodes have associated attributes, the privacy risks increase. 
%However, such risks have not been empirically evaluated. 
In this paper we quantitatively study the impact of binary node attributes on node privacy by employing machine-learning-based re-identification attacks and exploring the interplay between graph topology and attribute placement. 
Our experiments show that the population's diversity on the binary attribute consistently degrades anonymity.  
}
\end{abstract}
  \keywords{Node re-identification, machine learning attack, labeled networks.}
  
  \maketitle
%  \classification[PACS]{}
 % \communicated{...}
 % \dedication{...}

%   \journalname{Proceedings on Privacy Enhancing Technologies}
% \DOI{Editor to enter DOI}
%   \startpage{1}
%   \received{..}
%   \revised{..}
%   \accepted{..}

%   \journalyear{..}
%   \journalvolume{..}
%   \journalissue{..}

\section{Introduction}

%%% Alternative storyline:
%%% anonymization important
%%% utility important
%%% thus, dK relevant
%%% but how well does it work
%%% ML-based techniques in this era
%%% intuitively, 1K>2K in anonymity
%%% but is it always the case?
%%% so, we look at strength of the attacker not in terms of quantity but quality of information
%%% and the particularities of the dataset to anonymize

Real graph datasets are fundamental to understanding a variety of phenomena, such as epidemics, adoption of behavior, crowd management and political uprisings. %\anda{need good examples here}. 
At the same time, many such datasets capturing computer-mediated social interactions are recorded nowadays by individual researchers or by organizations. %, some for profit, others governmental. 
However, while the need for real social graphs and the supply of such datasets are well established, the flow of data from data owners to researchers is significantly hampered by serious privacy risks: even when humans' identities are removed, studies have proven repeatedly that de-anonymization is doable with high success rate~\cite{narayanan2011link, srivatsa2012deanonymizing, ji2014structure, korula2014efficient}. 
Such de-anonymization techniques reconstruct user identities using third-party public data and the graph structure of the naively anonymized social network: specifically, the information about one's social ties, even without the particularities of the individual nodes, is sufficient to re-identify individuals. 

%The accepted approach now is thus to anonymize social graphs by modifying the graph structure enough as to decouple the particular node identity from its social ties, yet preserving the graph characteristics in aggregate. 
%Various solutions have been proposed, some based on rewiring the original graph structure, others based on clustering, and others based on generating graphs from a graph signature. For all structural graph anonymization techniques, however, the challenge is the tension between providing privacy in the altered graph structure and preserving the accuracy of the structural characteristics of the original graph in the altered graph, which is what matters for their utility for research~\cite{aggarwal2011hardness}. 

Many anonymization methods have been proposed to mitigate the privacy invasion of individuals from the public release of graph data~\cite{ji2016survey}.
Naive anonymization schemes employ methods to scrub identities of nodes without modifying the graph structure.
Structural anonymization methods change the topology of the original graph while attempting to preserve (at least some of) the original graph characteristics~\cite{liu2008towards,
%hay2008resisting,
sala2011sharing,liu2016linkmirage}. Often the utility of an anonymized graph depends not only on preserving essential graph properties of the original graph, but also node attributes such as labels that identify nodes as cheaters or noncheaters in online gaming platforms~\cite{blackburn2014}.  

However, the effects of node attributes on the risks of re-identifications are not yet well understood. 
While intuitively any extra piece of information can be a danger to privacy, a rigorous understanding of what topological and attribute properties affect the re-identification risks is needed. 
In cases such as information dissemination, node attributes may be informed by the local graph topology.
\emph{How does the interplay between topology and node attributes affect node privacy?}

Our work assesses the additional vulnerability to re-identification attacks posed by the attributes of a labeled graph.  
We consider exactly one binary attribute to understand the lower bound of the damage that node attributes inflict.  
We focus our empirical study on the interplay between topology and labeling as a leverage point for re-identification. 
Because our focus is to understand in which conditions node re-identification is feasible, this study is independent of any anonymization technique. 
We apply machine learning techniques that use both topological and attribute information to re-identify nodes based on a common threat model.
% given access to an auxiliary graph with identifying information.  
Our study involves real-world graphs and synthetic graphs in which we control how labels are placed relative to ties to mimic the ubiquitous phenomena of homophily found in social graphs~\cite{mcpherson2001}.

Our empirical results show that the vulnerability to node re-identification depends on the population diversity with respect to the attribute considered. 
Using information about the distribution of labels in a node's neighborhood provides additional leverage for the re-identification process, even when labels are rudimentary.
Furthermore, we quantify the relative importance of attribute-related and topological features in graphs of different characteristics.

% why data release, and why anonymity
%Social graphs serve as mathematical models of network structures that represent interactions between real-world entities.  
%Such graphs are often mined to uncover insights about the structure and function of the interactions represented.  
%This substantial scientific value to the research community comes with risks: the release of such data, even with all personally identifying information (PII) removed, may jeopardize the privacy of individuals~\cite{backstrom2007wherefore}. 
%The Web-strip~\cite{web-strip-scandal} and AOL~\cite{AOLscandal} scandals are textbook examples on breaching the privacy of individuals by publicly releasing unsanitized data.

%\shnote{Has to say that we deal with non-PII information here in our work}

The remainder of this paper is organized as follows. 
Section~\ref{sec:related_work} outlines the related work. 
The system model to quantify anonymity is presented in Section~\ref{sec:methodology}. 
Section~\ref{sec:datasets} describes the characteristics of the datasets we used in our empirical investigations. 
we present our results in Section~\ref{sec:experiments} and discuss our contributions in Section~\ref{sec:discussions}.

\section{Related Work}
\label{sec:related_work}
% About dK-private graphs
%In this section, we analyze existing techniques to quantify the de-anonymizability of an anonymized graph. 
%While much has been published on quantifying the de-anonymizability of an anonymized graph, in this %section we limit our scope to the generation of anonymized graphs based on dK-graph models. 

%\ainote{one paragraph on dK-graphs to build intuition.}

% Taxonomy to de-anonymization attacks, and general quantification metric
The availability of auxiliary data helps reveal the true identities of anonymized individuals, as proven empirically in large privacy violation incidents~\cite{NetflixScandal,griffith2005messin}.
%For example, Griffith and Jakobsson used public records such as marriage and birth information to derive a mother's maiden name~\cite{griffith2005messin}
Similarly, in the case of graph de-anonymization attacks, information from an auxiliary graph is used to re-identify the nodes in an anonymized graph~\cite{narayanan2009anonymizing}. 
The quality of such an attack is determined by the rate of correct re-identification of the original nodes in the network.
In general, de-anonymization attacks harness structural characteristics of nodes that uniquely distinguish them~\cite{ji2016survey}.
Many such attacks can be categorized into \emph{seed-based} and \emph{seed-free}, based on the prior seed knowledge available to an attacker~\cite{ji2016survey}. 

% seed based and seed free attacks
In seed-based attacks, sybil nodes~\cite{backstrom2007wherefore} or some known mappings of nodes in an auxiliary graph aid the re-identification of anonymized nodes~\cite {narayanan2011link, srivatsa2012deanonymizing, ji2014structure, ji2016general, korula2014efficient}.
%In seed-based attacks, the process of de-anonymization is conducted to re-identify nodes and ties with the support of sybil nodes~\cite{backstrom2007wherefore} or some known mappings of nodes in an auxiliary graph~\cite{narayanan2011link, srivatsa2012deanonymizing, ji2014structure, ji2016general, korula2014efficient}.
%\cgnote{Suggested Edit: In seed-based attacks, sybil nodes~\cite{backstrom2007wherefore} or some known mappings of nodes in an auxiliary graph aid the re-identification of anonymized nodes~\cite {narayanan2011link, srivatsa2012deanonymizing, ji2014structure, ji2016general, korula2014efficient}.}
The effectiveness of such attacks is influenced by the quality of the seeds~\cite{Sharad2016benchmark}.

In seed-free attacks, the problem of deanonymization is usually modeled as a graph matching problem. 
Several research efforts have proposed statistical models for the re-identification of nodes without relying on seeds, such as the Bayesian model~\cite{pedarsani2013bayesian} or optimization models~\cite{ji2014structural, ji2016structuralsf}. 
Many heuristics are used in the propagation process of re-identification, exploiting graph characteristics such as degree~\cite{gulyas2016efficient}, k-hop neighborhood~\cite{yartseva2013performance}, linkage-covariance~\cite{aggarwal2011hardness}, eccentricity~\cite{narayanan2009anonymizing}, or community~\cite{nilizadeh2014community}.

Recently, there have been efforts to incorporate node attribute information into deanonymization attacks.
Gong et al.~\cite{gong2014joint} evaluate the combination of structural and attribute information on link prediction models.
Attributes not present may be inferred through prior knowledge and network homophily.
% Their initial focus was determining the pattern of evolution in dynamic networks, but node-attribute link prediction models have been expanded to bolster deanonymization attacks.
Qian et al.~\cite{qian2016anonymizing} apply link prediction and attribute inference to deanonymization by quantifying the prior background information of an attacker using knowledge graphs. In knowledge graphs, edges not only represent links between nodes but also node-attribute links and link relationships among attributes. 
% Each link is assigned a confidence score based on the probability of structural or attribute similarity.
% The knowledge graph serves the function of the auxiliary allowing both topological and attribute mappings to improve the deanonymization process. 
% Link deanonymization proceeds by BFS and linear regression following the highest confidence score.
The deanonymization attack in~\cite{JiMittal2016De-SAG} maps node-attribute links between an anonymized graph and its auxiliary. 
In addition to structural similarity, nodes are matched by attribute difference, the union of the attributes of the node in the anonymized and auxiliary divided by their intersection.

% why such ad-hoc metric not good, moving into Sharad
However, the success rate of a de-anonymization process is often reported in the literature as dependent on the chosen heuristic of the attack, which is typically designed with knowledge of the anonymization technique.
Comparing the strengths of different anonymization techniques thus becomes challenging, if not impossible. 
Recently, Sharad~\cite{Sharad2016benchmark} proposed a general threat model to measure the quality of a deanonymization attack which is independent of the anonymization scheme. 
He proposed a machine learning framework to benchmark perturbation-based graph anonymization schemes.
This framework explores the hidden invariants and similarities to re-identify nodes in the anonymized graphs~\cite{sharad2013anonymizing, sharad2014automated}.
Importantly, this framework can be easily tuned to model various types of attacks. % in the absence of seed knowledge. % available for the mapping between anonymized and auxiliary graphs.
 
Several researchers propose theoretical frameworks to examine how vulnerable or deanonymizable any (anonymized) graph dataset is, given its structure~\cite{pedarsani2011privacy,ji2014structural, ji2015your, ji2016relative}.
However, some techniques are based on Erd{\"o}s-R{\`e}nyi (ER) models~\cite{pedarsani2011privacy}, while others make impractical assumptions about the seed knowledge~\cite{ji2015your}.
% dense seeds are available
Ji et al.~\cite{ji2016relative} also introduced a configuration model to quantify the deanonymizablity of graph datasets by considering the topological importance of nodes.
The same set of authors analyzed the impact of attributes on graph data anonymity~\cite{JiMittal2016De-SAG}. 
They show a significant loss of anonymity when more node-attribute relations are shared between anonymized and auxiliary graph data.
Specifically, they measure the entropy present in node-attribute mappings available for an attacker.
As the entropy decreases, the graph loses node anonymity.
% Such observations are drawn under a randomized model of attribute attachment.

The main aspects distinguishing this study from existing works are as follows: i) In our work, 
%While they analyze the benefits of $dK$ graph generators through models of DP,
we study the inherent conditions in graphs that provide resistance/vulnerability to a general node re-identification attack based on machine learning techniques.
ii) To the best of our knowledge, this is the first work that quantifies the privacy impact of node attributes under an attribute attachment model biased towards homophily.
iii) We analyze the interplay between the intrinsic vulnerability of the graph structure and attribute information.

\section{Methodology}
\label{sec:methodology}

%\shnote{More importantly, all the existing de-anonymizability analysis for graph data are structure- based. Fundamentally understanding the impacts of user- associated attributes on the anonymity of graph data is still an open problem.} \cite{JiMittal2016De-SAG}

%\shnote{what are the impacts of the attribute information on the anonymity/de-anonymizability of graph data?}\cite{JiMittal2016De-SAG}
Our main objective is to quantitatively estimate the vulnerability to re-identification attacks added by node attributes. 
In particular, we ask: \emph{Given a graph topology, how much better does a  
node re-identification attack perform when the node attributes are included in the attack compared to when there is no node attribute information available to the attacker?}

We are interested in measuring the intrinsic vulnerability of a graph with attributes on nodes, in the absence of any particular anonymization technique on topology or node attributes. 
The intuition is that particular graphs are inherently more private: for example, in a regular graph, nodes are structurally indistinguishable. 
Adding attributes to nodes, however, may contribute extra information that could make the re-identification attack more successful. 
Consider another example, in a highly disassortative network (such as a sexual relationships network), knowing the attribute values (i.e., gender) of a few nodes will quickly lead to correctly inferring the attribute values of the majority of nodes, and thus possibly contributing to the re-identification of more nodes.  
Thus, the questions we address in this study also include: \emph{How does the distribution of node attributes affect the intrinsic vulnerability to a re-identification attack of a labeled graph topology?}  
%\ainote{what I'm trying to explain here is the attack on the original graph, not an anonymized version. I don't know if we've seen it in the literature before.}

To answer these question, we developed a machine learning-based re-identification attack inspired from that presented in~\cite{Sharad2016benchmark}. 
We use the same threat model (Section~\ref{sec:threat-model}) that aims at finding a bijective mapping between nodes in two different
% already anonymized \ainote{what do we mean by already anonymized? In this case, we only have the node identities removed} 
graphs. 
We mount a machine-learning based attack (Section~\ref{sec:machine-learning-attack}), in which the algorithm learns the correct mapping between some pairs of nodes from the two graphs, and estimates the mapping of the rest of the dataset. 
As input data, we use both real and synthetic datasets (as presented in Section~\ref{sec:datasets}).  
\subsection{The Threat Model}
\label{sec:threat-model}
%-------------------------------------
%More often, the problem of re-identification of nodes in an anonymized graph is mapped to the identification of corresponding nodes in an 
%auxiliary graph.
%This is further being reduced to the variations of \textit{graph isomorphism problem} \cite{corneil1970efficient}.
%%We follow the general threat model used in~\cite{Sharad2016benchmark} to measure the success of a de-anonymization attack.
The threat model we consider is the classical threat model in this context~\cite{pedarsani2011privacy}: 
The attacker aims to match nodes from two networks whose edge sets are correlated. 
% \ainote{Why unlabeled? Why structurally similar? The threat model works without these assumptions, no?}
% \shnote{In Pedrsani model, the goal is to match the vertices of two unlabeled graphs whose edge sets are correlated but not necessarily equal (structurally similar)}
% \shnote{This is equivalent to the assumption that the two graphs can have different vertex label sets in the unseen examples.(unlabel)}
We assume each node is associated with a binary valued attribute, and this attribute is publicly available. 
Common examples of such attributes are gender, professional level (i.e., junior or senior), or education level (i.e., higher education or not). 

For clarity, consider the following example: an attacker has access to two networks of individuals in an organization that represent the communication patterns (e.g., email) and friendship information available from an online social network. 
Individuals in the communication network are described by professional seniority (e.g., junior or senior), while individuals in the friendship network are described by gender.
%%More generally, an individual might have different identities in two domains, having non-identical but correlated social neighborhood structure~\cite{pedarsani2011privacy}.
% Both networks have been anonymized by removing \ainote{removing, really? Or possibly perturbing it?} the original node labels and by perturbing edges.
% Anonymization schemes may also alter existing attribute information associated to nodes.
% However, we assume this process maintain the same distribution of attribute values.
These graphs are structurally overlapping, in that some individuals are present in both graphs, even if their identities have been removed. 
% attributes have been altered, while their links have possibly been perturbed.  \ainote{why are we talking about altered attributes and perturbed edges?!}
The attacker's task is to find a bijective mapping between the two subsets of nodes in the two graphs that correspond to the individuals present in both networks.

%-------------------------------------
\subsection{Machine Learning Attack}
\label{sec:machine-learning-attack}
%-------------------------------------

\begin{figure}[t]
	\centering
	\includegraphics[width=1\linewidth]{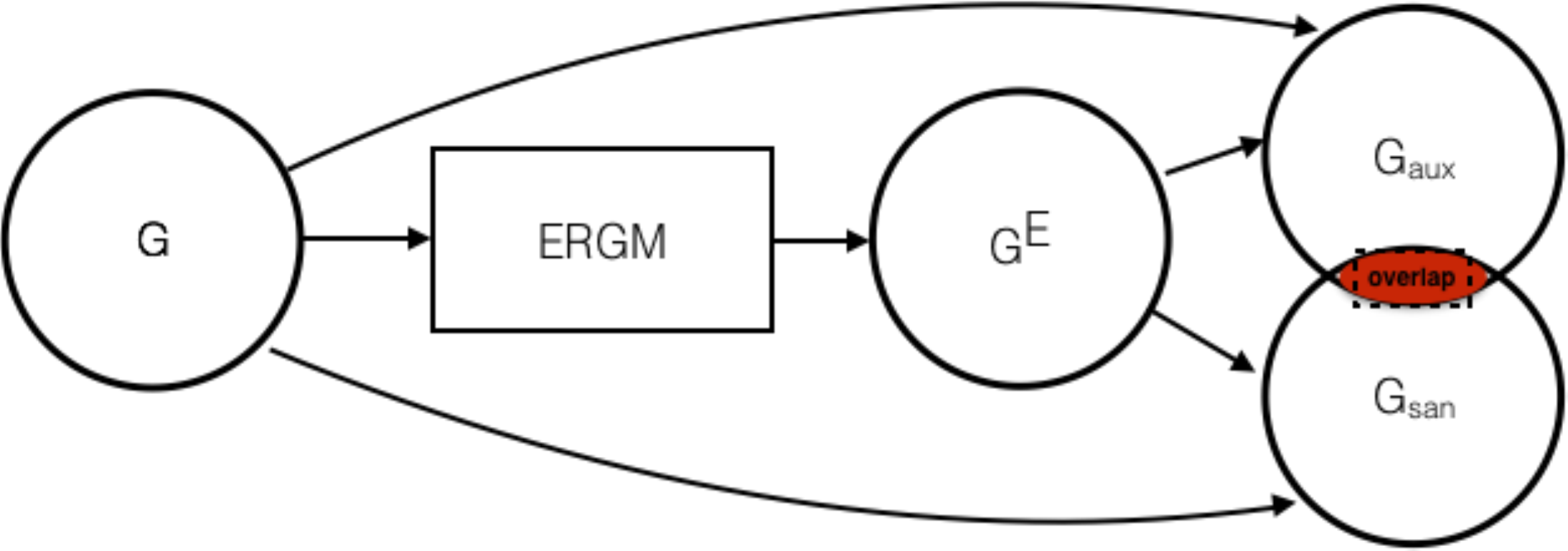}
	\caption{Graph mining system flow to generate node pairs with the ground truth of identical and non-identical pairs. %\ainote{I don't understand this:}The instances of $G_{aux}$ and $G_{san}$ pairs approximate the auxiliary and anonymized graphs at attacker's disposal.\shnote{This is being explained in the text, just a recall, but we can remove it from the caption}
	}
	\label{fig:graphminingflow}
\end{figure}

We assume that the adversary has a sanitized graph $G_{san}$ that could be associated with an auxiliary graph $G_{aux}$ for the re-identification attack (as depicted in Figure~\ref{fig:graphminingflow}). 
As in the scenario discussed above, $G_{san}$ could be the communication network, while $G_{aux}$ is the friendship network of a set of individuals in an organization.

In order to model this scenario using real data, we split a real dataset graph $G=(V,E)$ into two subgraphs $G_1=(V_1,E_1)$ and $G_2=(V_2,E_2)$, such that $V_1 \subset V$, $V_2 \subset V$ and $V_1 \cap V_2 = V_\alpha$, where $V_\alpha \ne \phi$. 
The fraction of the overlap $\alpha$ is measured by the Jaccard coefficient of two subsets: $\alpha=\frac{|V_1 \cap V_2|}{|V_1 \cup V_2|}$. 
In the shared subgraph induced by the nodes in $V_\alpha$, nodes will preserve their edges with nodes from $V_\alpha$ but might have different edges to nodes that are part of $V_1 - V_{\alpha}$ or part of $V_2 - V_{\alpha}$.
Each nodes $v \in V_1 \cup V_2$ maintains its original attribute value. 

In an optimistic scenario, an attacker has access to a part of the original graph (e.g., $G_1$) as auxiliary data and to an unperturbed subgraph (e.g., $G_2$) as the sanitized data whose nodes the attacker wants to re-identify.
We use $G_1$ and $G_2$ as baseline graphs to measure the impact of attributes on de-anonymizability of network data.
It is also possible to split $G_1$ and $G_2$ recursively into multiple overlapping graphs, maintaining the same values of overlap parameters as above. 
This allows us to assess the feasibility of the de-anonymization process for large networks by significantly reducing the size of $G_1$ and $G_2$.

% However, this is unrealistic, since researchers and practitioners are well aware now that without perturbing the structure of the graph, even with identifiable node and edge attributes removed, there is no anonymity~\cite{narayanan2009anonymizing}.
% Hence, in order to reproduce a realistic scenario, we anonymize both $G_1$ and $G_2$ graphs (as depicted in Figure~\ref{fig:dkminingflow}).
% Note that this is a more challenging scenario for the attacker.
% The (not anonymized) graphs $G_1$ and $G_2$ will be used as a baseline to measure the anonymization power of the $dK$-based techniques. 

% We also produced $m$ number of anonymized graphs, in order to increase the dataset used for experimentation. 
% The anonymized graphs are then split again into two overlapping graphs, maintaining the same values of overlap parameters as above.
The resulting graphs are now the equivalent of the email/friendship networks we used as an example above. 
The overlap is the knowledge repository that the attacker uses for de-anonymization~\cite{henderson2011s}. 
Part of this knowledge will be made available to the machine learning algorithms. 

Previous work shows that the larger $\alpha$, the more successful the attack. %~\cite{Sharad2016benchmark}.
However, the relative success of attacks under different anonymization schemes is observed to be independent of $\alpha$~\cite{Sharad2016benchmark}.
In order to experiment with a homogeneous attack, we set the value of $\alpha=0.2$, and we build $V_\alpha$ by building a breadth-first-search tree starting from the highest degree node (BFS-HD) in $G$.
While other alternatives are certainly possible, we chose this approach for two reasons.
First, it appears that the threat model we used is quite sensitive to the sampling process when generating $G_1$ and $G_2$~\cite{pedarsani2011privacy}. 
To avoid sampling bias, we chose a BFS-HD split to have a deterministic set of nodes in $V_\alpha$.
% based on a uniformly distributed overlap. 
% \ainote{I don't understand what a uniformly distributed overlap means. Also, I wonder what you're trying to say here. Is it that you want a deterministic set of nodes in $V_\alpha$, thus start BFS (deterministic process) from a deterministically identified node?}
% \shnote{Yes.Updated.}
Second, we empirically found that BFS-HD provides the maximally informed seeds for an adversary to propagate the re-identification process, thus providing a best-case scenario for the attacker.

% in four different ways: i)~as a random collection of nodes from the original graph $G$ (R); 
% ii)~by selecting the highest degree nodes from $G$ (HD); iii)~ by building a breadth-first-search tree starting from a randomly selected node in $G$ (BFS-R); and iv)~by building a breadth-first-search tree starting from the highest degree node in $G$ (BFS-HD). 

%.......................................
\subsubsection{\textbf{Node Signatures}}
\label{sec:node-signatures}
%.......................................

\begin{figure}[!t]
	\centering
	\includegraphics[width=0.75\linewidth]{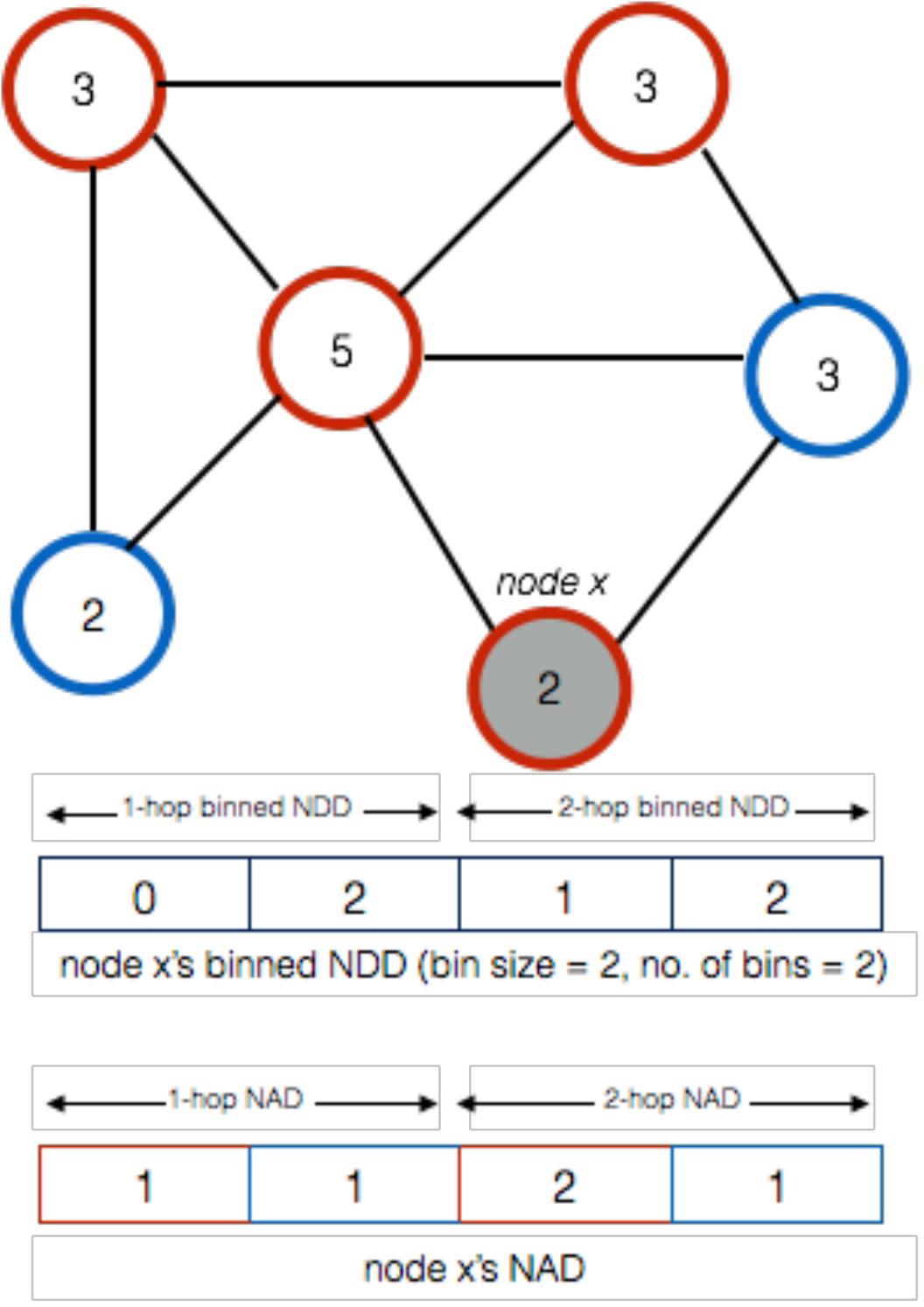}
	\caption{Example of a node signature defined as a combined feature vector made up from NDD and NAD vectors. In the NDD vector, each bin value corresponds to the number of nodes that have a degree value represented in the bin range, such that the $j^{th}$ bin holds the nodes in the degree ($k_j$) range $j\times b \leq k_j < (j+1)\times b$. If the degree exceeds the maximum range, such nodes are included in the last bin. Further, both 1-hop and 2-hop NDDs are calculated and merged. For example, node $x$ has no 1-hop neighbor nodes that have degree in the range of $1-2$, and one 2-hop neighbor node that its degree is in the range of $1-2$. In the NAD vector, each element corresponds to the number of nodes with the given attribute. Both 1-hop and 2-hop NADs are calculated and merged. Node $x$ has one 1-hop neighbor node, and two 2-hop neighbor nodes with the attribute \texttt{Red}. Note that the node value represents the associated degree, and the border color represents the node attribute \texttt{Red} or \texttt{Blue}.
	}
	\label{fig:nodesignature}
\end{figure}

Since we are employing machine learning techniques to re-identify nodes in a graph, we need to represent nodes as feature vectors. 
We define the node $u$'s features using a combination of two vectors made up from its neighborhood degree distribution (NDD) and neighborhood attribute distribution (NAD) (as depicted in Figure \ref{fig:nodesignature}). 

NDD is a vector of positive integers where $NDD^q_u[k]$ represents the number of $u$'s neighbors at distance $q$ with degree $k$.
%\ainote{I think we can improve the notation -- in our case k is really a bin, correct? And since it's a vector, we want to define the i-th element of the vector in a sound way. To rewrite this part -- we can discuss. Also, the caption is too long and I am not sure the notation/presentation in the caption is consistent with the text in here. To revise that as well.}
We concatenate the binned version of $NDD^1_u$ with the binned version of $NDD^2_u$ to define the node $u$'s NDD signature.
% We employ a binning strategy to capture the aggregate structure of ego networks which is expected to be robust against edge perturbation due to anonymization~\cite{sharad2016learning}.
% Further, we employ a binning strategy 
% Thus, we represent the signature of a node $u$ as the concatenation of the binned version of $NDD^1_u$ with the binned version of $NDD^2_u$. 
We use a bin size of $50$, which was shown empirically~\cite{Sharad2016benchmark} to capture the high degree variations of large social graphs.
For each $q$, we use $21$ bins, which would correspond to a larger node degree of $1050$. All larger values are binned in the last bin.
This binning strategy is designed to capture the aggregate structure of ego networks, which is expected to be robust against edge perturbation~\cite{sharad2016learning}.
% The decision of using NDD is supported by the empirical observation that a large proportion of nodes in a graph are uniquely distinguishable by their NDD~\cite{Sharad2016benchmark}.

NAD is defined by $NAD^q_u[i]$ which represents the number of $u$'s neighbors at distance $q$ with an attribute value $i$. It is shown experimentally that the use of neighbor attributes as features often improves the accuracy of edge classification tasks~\cite{mcdowell2013labels}.

We use the notation GS to represent the prediction results from the input features made up from the topology (e.g., NDD). GS(LBL) to represent features from both the topology and attribute information (e.g., concatenation of NDD and NAD vectors).
% \cgnote{The preceding might be clearer as two sentences.}
% \shnote{Fixed.}

%.......................................
\subsubsection{\textbf{Random Forest Classification}}
\label{sec:random-forest-classification}
%.......................................

Note that the nodes in $G_{san} \cap G_{aux}$, common to both graphs, can be recognized as being the same node (identical) in the two graphs based on their node identifier. 
Non-identical nodes are unique to each $G_{san}$ and $G_{aux}$ and would not exist in the overlap.
In the classification task, we wish to output 1 for an identical node pair and 0 for a non-identical node pair.
This is the ground truth against which we measure the accuracy of the learning algorithms.

% \ainote{The ground truth is also made of the nodes that are not in the overlap, right? That would be the ground truth for pairs of nodes that are not identical. If this makes sense, then we need to clarify.}
% \shnote{Updated.}

We generate examples for the training phase of the deanonymization attack by randomly picking node pairs from the sanitized $(G_{san})$ and the auxiliary $(G_{aux})$ graphs, respectively. 
In most cases, we have an unbalanced dataset with the degree of imbalance depending on the overlap parameter $\alpha$, where the majority is non-identical node pairs.
We use the reservoir sampling technique~\cite{haas2016data} to take $\ell=1000$ balance sub-samples from the population $S$, and the SMOTE algorithm~\cite{smote} as an over-sampling technique for each sub-sample.
%Each sample is split into training and testing sets. 
Each sample is trained by a forest of 100 random decision trees that allows the algorithm to learn features.
% Accuracy is measured on unseen examples from the same sample space.
Gini-index is used as an impurity measure for the random forest classification.
Given the size $\alpha$ of the overlap,
% \ainote{do we still have a choice of the overlap?:} 
we measure the quality of the classifier on the task of differentiating two nodes as identical or not. 
% We thus use $2 \times m \times \ell \times 100$ decision trees per $dK$-space for the learning process in each graph space.
% Table~\ref{tbl:sample_space} summarizes the statistics of the sample population.

%.......................................
\subsubsection{Metrics}
\label{sec:metrics}
%.......................................

We measure the accuracy of the classifier in determining whether a randomly chosen pair of nodes (with one node in $G_{san}$ and another in $G_{aux}$) are identical or not.
% Accuracy is measured by the mean value of F1-score, which is calculated over a distribution of $5 \times 2$ cross-validation scores.
We use F1-score to evaluate the quality of the classifier. 
F1-score is the harmonic mean between precision and recall, typical metrics for prediction output of machine learning algorithms. 
% Recall that, we have $\ell$ number of balanced samples of node pairs per graph space.

For each data sample, we perform $5\times2$ cross-validation to evaluate the classifier and record the mean F1-score.
We thus build two vectors of mean F1-scores, each of size $\ell=1000$ (as described above), one for the labeled (GS(LBL)) and one for the unlabeled network topology (GS). 
%The output would be two vectors of such accuracy values, where each vector include mean F1-scores to describe the quality of classifiers under the respective set of input features: GS and GS(LBL).
An important aspect of these vectors is that they are related in the sense that the $i^{th}$ element in one vector represents the same sample as the $i^{th}$ element of the other vector.
This is important for the pairwise comparison of the two mean F1-score vectors. 

We perform a standard T-test on these two vectors and report the T-statistic value.
The T-statistic value is a measure of how close to the hypothesis an estimated value is. 
In our case, the hypothesis is the prediction accuracy of the node identities in the unlabeled graph (GS) and the estimated value is the prediction accuracy in the labeled graph (GS(LBL)). 
Thus, a large T-statistic value implies a significantly better prediction accuracy of node identities in GS(LBL) than in GS. 
In such cases, we can say that the network with node attributes is more vulnerable to node re-identification. % when T-statistic is positive and large.
This value serves as our statistical measurement to quantify the vulnerability cost of node attributes.

\section{Datasets}
\label{sec:datasets}

Because our work is empirically driven, a larger set of test datasets promises a better understanding of the relations between vulnerability to re-identification attacks and the particular characteristics of the node attributes (such as fractions of attributes of a particular value or the assignment of attributes to topologically related nodes). 
In this respect, real datasets are always preferable to synthetic ones, as they potentially encapsulate phenomena that are missing in the graph generative models. 
As an example, until very recently, the relation between the local degree assortativity coefficient and node degree was not captured in graph topology generators~\cite{sendina2016assortativity}. %~\cite{Mussmann2015assortativity}. 
%We selected five real datasets described in Section~\ref{sec:realdatasets}. 

However, relying only on real datasets has its limitations, due to the scarcity of relevant data (in this case, networks with binary node attributes) and the difficulty of covering the relevant space of graph metrics when relying only on available real datasets. 
Thus, in this work, we combine real networks (described in Section~\ref{sec:realdatasets}) with synthetic networks generated from the real datasets. 
For generating synthetic labelled networks, we employ ERGMs~\cite{holland1981exponential,wasserman1996logit} and a controlled node-labeling algorithm as described in Section~\ref{sec:synthetic_graphs}. 
%To add synthetic labeling following controlled processes, we use a synthetic labeling technique introduced in~\cite{Skvoretz2013} and described in Section~\ref{sec:labeling}. 

%-------------------------------------
\subsection{Real Network Datasets}
\label{sec:realdatasets}
%-------------------------------------

We chose six publicly available datasets from four different contexts and generated eight networks with binary node attributes.

\begin{itemize}
% We chose five publicly available network datasets. 
%Each network has nodes associated with a binary attribute.
% is a bi-modal projection on a binary attribute.
% so that the effect of the deanonymization process on labeled data could be analyzed.
\item \texttt{polblogs}~\cite{adamic2005political} is an interaction network between political blogs during the lead up to the 2004 US presidential election. 
This dataset includes ground-truth labels identifying each blog as either conservative or liberal. 

\item \texttt{fb-dartmouth}, \texttt{fb-michigan}, and \texttt{fb-caltech}~\cite{traud2012social} are Facebook social networks extant at three US universities in 2005.
A number of node attributes such as dorm, gender, graduation year, and academic major are available. 
We chose two such attributes that could be represented as binary attributes: gender and occupation, whereby occupation we could identify the attribute values ``student'' and ``faculty''. 
From each dataset, we obtained two networks with the same topology but different node attribute distributions. 

\item \texttt{pokec-1}~\cite{takac2012data} is a sample of an online social network in Slovakia. While the Facebook samples are university networks, Pokec is a general social platform whose membership comprises 30\% of the Slovakian population. \texttt{pokec-1} is a one-fortieth sample. 
%\ainote{Can we say more such that we show that is a different context than Facebook? Maybe it included x\% of the population in an European country, thus it is not confined to an American academic environment?} \ainote{What sampling method was used and why did we choose that sampling technique?}
This dataset has gender information available as a node attribute.

\item \texttt{amazon-products}~\cite{leskovec2007dynamics} is a bi-modal projection of categories in an Amazon product co-purchase network. 
Nodes are labeled as ``book'' or ``music'', edges signify that the two items were purchased together.  
%The two product groups represented as binary attribute values are books and music. 
% There is an edge between products if they were purchased together. The last two datasets are archived by the Stanford Network Analysis Project.
\end{itemize}

As Table~\ref{tbl:dataset} shows, the networks generated from these datasets have different graph characteristics. 
For example, the density ($d$) of the graphs varies across three orders of magnitude, while degree assortativity oscillates between disassortative (for \texttt{polblogs}, $r = -0.22$, where there are more interactions between popular and obscure blogs than expected by chance) to assortative (as expected for social networks). All topologies except for \texttt{amazon-products} have small average path length. 

The metrics $p$ and $\tau$ shown in Table~\ref{tbl:dataset} are inspired from the synthetic node labeling algorithm used for generating synthetic graphs (and presented later), and they also show high variation across different networks. 
Intuitively, $p$ captures the diversity of attribute values in the node population (with $p=0.5$ showing equal representation of the attributes) while $\tau$ captures the homophily phenomenon (that functions as an attraction force between nodes with identical attribute values). 
The homophilic attraction metric $\tau$ varies between 0 in \texttt{pokec-1} (thus, no higher than chance preference for social ties with people of the same gender in Slovakia) to 0.99 in \texttt{amazon-products} (books are purchased together with other books much more strongly than given by chance).  
The diversity metric $p$ varies between the overrepresentation of males in the US academic Facebook networks (8\% female representation) to an almost perfect political representation in the \texttt{polblogs} dataset (where $p=0.48$). 
Note that, we only consider $p$ as the minimum proportion of two node groups due to the symmetric nature of attributes in our experiments.
% Note that the maximum value for $p$ is 0.5.  \ainote{conflict with later claim.}

This wide variation in graph metrics values is what motivated our choice for these set of real networks. 
We opted to include the three Facebook networks from similar contexts to also capture more subtle variations in network characteristics.

% %The dk graph generators are given the respective degree distributions and other graph characteristics (such as clustering coefficient) of a set of real networks. 
% We chose five publicly available datasets that represent real social networks of various types. 
% \texttt{fb107}~\cite{leskovec2012learning} represents social circles of an ego in Facebook. 
% %%\ainote{huh? it's only the contacts of one person + their contacts? Why is this a relevant network?}.
% %%\shnote{This would give us good points on comparison when we talk about degree one nodes in the experiments, because ego network lack such nodes}
% \texttt{caGrQc}~\cite{leskovec2007graph} is a co-authorship network between the authors of papers submitted to the scientific domain of general relativity and quantum cosmology.
% %\ainote{in what context and domain? Who are the nodes, what are the edges?}.
% \texttt{soc-anybeat}~\cite{Fire2012} is an interaction network available in the Anybeat online community, which is a public gathering place across the world.
% \texttt{soc-gplus}~\cite{networkrepo} is a follower network from Google+.
% Finally, \texttt{web-frwikinews-user-edits}~\cite{networkrepo} is a discussion network of Wikipedia pages. 
% Table~\ref{tbl:dataset} summarizes the graph statistics of these datasets.

\begin{table*}[!t]
	\centering
	\caption{
		Graph properties of the real network datasets. All graphs are undirected, and nodes are annotated with a binary valued attribute. E.g., nodes in the polblogs network have the attribute \texttt{party} with values; conservative and liberal. For simplicity, binary values are presented using the notation of \texttt{R} and \texttt{B}, together with the distributions of such values over nodes and edges. $p$ and $\tau$ present the estimated parameter values of the attraction model. Density $(\bar{d})$ is the fraction of all possible edges, transitivity $(C)$ is the fraction of triangles of all possible triangle in the network. degree-assortativity $(r)$ measures the similarity of relations depending on the associated node degree. Average path length $(\kappa)$ depicts the average shortest path length between any pairs of nodes.
	}
	\label{tbl:dataset}
% 	\scalebox{0.}{
	\begin{tabular}{|r|r|r|r|r|r|r|r|r|r|r|r|}
		\hline 
		\multirow{2}{*}{Network} & \multicolumn{2}{|c|}{$|N|$} & \multicolumn{3}{|c|}{$|E|$} & \multirow{2}{*}{$p$} & \multirow{2}{*}{$\tau$} & \multirow{2}{*}{$\bar{d}$} & \multirow{2}{*}{$C$} & \multirow{2}{*}{$r$} &  \multirow{2}{*}{$\kappa$} \\ 
		\cline{2-6}
		& $R (\%)$ &  $B (\%)$ & $R - R (\%)$ &  $B - B (\%)$ & $R - B (\%)$  &  &   & & & & \\ 
		\hline

			polblogs & \multicolumn{2}{|c|}{$1224$} & \multicolumn{3}{|c|}{$16718$} & \multicolumn{2}{c|}{} & \multirow{2}{*}{$0.02$} & \multirow{2}{*}{$0.22$} & \multirow{2}{*}{$-0.22$} &  \multirow{2}{*}{$2.49$} \\ 
		\cline{1-8}
		(party) & $48$ &  $52$ & $44$ &  $48$ & $8$  &  $0.48$ & $0.84$  & & & & \\ 
		\hline \\  \hline

			fb-caltech & \multicolumn{2}{|c|}{$769$} & \multicolumn{3}{|c|}{$16656$} & \multicolumn{2}{c|}{} & \multirow{3}{*}{$0.05$} & \multirow{3}{*}{$0.29$} & \multirow{3}{*}{$-0.06$} &  \multirow{3}{*}{$1.33$} \\ 
		\cline{1-8}
		(gender) & $91.5$ &  $8.5$ & $92.8$ &  $0.2$ & $7$  &  $0.08$ & $0.52$  & & & & \\ 
		\cline{1-8}
		(occupation) & $72$ &  $28$ & $69$ &  $8$ & $23$  & $0.28$ & $0.42$  & & & & \\
		\hline \\  \hline

			fb-dartmouth & \multicolumn{2}{|c|}{$7694$} & \multicolumn{3}{|c|}{$304076$} &  \multicolumn{2}{c|}{} & \multirow{3}{*}{$0.01$} & \multirow{3}{*}{$0.15$} & \multirow{3}{*}{$0.04$} &  \multirow{3}{*}{$2.76$} \\ 
		\cline{1-8}
		(gender) & $86.5$ &  $13.5$ & $83.2$ &  $0.9$ & $15.9$  & $0.14$ &  $0.34$ & & & & \\ 
		\cline{1-8}
		(occupation) & $62$ &  $38$ & $58$ &  $18$ & $24$  & $0.38$ &  $0.5$ & & & & \\
		\hline \\  \hline
		
			fb-michigan & \multicolumn{2}{|c|}{$30147$} & \multicolumn{3}{|c|}{$1176516$} & \multicolumn{2}{c|}{} & \multirow{3}{*}{$0.0026$} & \multirow{3}{*}{$0.13$} & \multirow{3}{*}{$0.115$} &  \multirow{3}{*}{$3.05$} \\ 
		\cline{1-8}
		(gender) & $92.2$ &  $7.8$ & $90.5$ &  $0.2$ & $9.3$  &  $0.08$ & $0.37$  & & & & \\ 
			\cline{1-8}
		(occupation) & $77.5$ &  $22.5$ & $72$ &  $9$ & $19$  &  $0.22$ & $0.46$  & & & & \\ 
		\hline  \\  \hline

% 		gplus & \multicolumn{2}{|c|}{$105978$} & \multicolumn{3}{|c|}{$12214600$} & \multicolumn{2}{c|}{} & \multirow{3}{*}{$0.002175$} & \multirow{3}{*}{$-$} & \multirow{3}{*}{$-$} &  \multirow{3}{*}{$-$} \\ 
% 		\cline{1-8}
% 		(gender) & $52.5$ &  $47.5$ & $51.5$ &  $9.5$ & $39$  &  $0.47$ & $0.218$  & & & & \\ 
% 		\cline{1-8}
% 		(education) & $-$ &  $-$ & $-$ &  $-$ & $-$  &  $-$ & $-$  & & & & \\ 
% 		\hline  \\  \hline

			pokec-1 & \multicolumn{2}{|c|}{$265388$} & \multicolumn{3}{|c|}{$700352$} & \multirow{2}{*}{$0.46$} & \multirow{2}{*}{$0$} & \multirow{2}{*}{$2 \times 10^{-5}$} & \multirow{2}{*}{$0.0068$} & \multirow{2}{*}{$-0.044$} &  \multirow{2}{*}{$5.66$} \\ 
		\cline{2-6}
		(gender) & $46$ &  $54$ & $18.6$ &  $22.4$ & $59$  &  &   & & & & \\ 
		\hline   \\  \hline
		
%			\multirow{2}{*}{pokec-2(gender)} & \multicolumn{2}{|c|}{$277937$} & \multicolumn{3}{|c|}{$721106$} & \multirow{2}{*}{$0.44$} & \multirow{2}{*}{$0$} & \multirow{2}{*}{$2 \times 10^{-5}$} & \multirow{2}{*}{$0.018$} & \multirow{2}{*}{$-0.06$} &  \multirow{2}{*}{$\kappa$} \\ 
%		\cline{2-6}
%		& $44.4$ &  $55.6$ & $19$ &  $21.2$ & $59.8$  &  &   & & & & \\ 
%		\hline 

			amazon-products & \multicolumn{2}{|c|}{$303551$} & \multicolumn{3}{|c|}{$835326$} & \multirow{2}{*}{$0.18$} & \multirow{2}{*}{$0.99$} & \multirow{2}{*}{$1.8 \times 10^{-5}$} & \multirow{2}{*}{$0.21$} & \multirow{2}{*}{$-0.06$} &  \multirow{2}{*}{$17.42$} \\ 
		\cline{2-6}
		(category) & $82$ &  $18$ & $83.4$ &  $16.4$ & $0.2$  &  &   & & & & \\ 
		\hline 
		
	\end{tabular} 
% 	}
\end{table*}

%-------------------------------------
\subsection{Synthetic Graphs}
\label{sec:synthetic_graphs}
%-------------------------------------
In order to be able to control graph characteristics and node attribute distributions, we also generated a number of synthetic graphs comparable with the real datasets just described. 
The graph generation included two aspects: topology generation, for which we opted for ERGMs, and node attribute assignments, for which we implemented the technique proposed in~\cite{Skvoretz2013}. 

%.......................................
\subsubsection{Varying Topology via ERGMs}
\label{sec:ERGM}
%.......................................

Exponential-family random graph models (ERGMs) or p-star models~\cite{holland1981exponential,wasserman1996logit} are used in social network analysis for stipulating, within a set structural parameters, distribution probabilities for networks. 
Its primary use is to describe structural and local forces that shape the general topology of a network. 
This is achieved by using a selected set of parameters that encompass different structural forces (e.g., homophily, degree correlation/assortativity, clustering, and average path length). 
Once the model has converged, we can obtain maximum-likelihood estimates, model comparison and goodness-of-fit tests, and generate simulated networks tied to the relationship between the original network and the probability distribution provided by the ERGM.

Our interest in ERGMs is based on simulating graphs that retain set structural information from the original graph to generate a diverse set of graph structures.
We used R~\cite{team2014r} and the \texttt{statnet} suite~\cite{handcock2014statnet}, which contains several packages for network analysis, to produce ERGMs and simulate graphs from our real-world network datasets.
In this case, we focused on three structural aspects of the graphs: clustering coefficient, average path length, and degree correlation/assortativity. 
% \shnote{What supports our choices? see the comments.}
% assortativity often distinguishes between social and technological networks while clustering and average path length are the defining features of small world networks 

For the ERGM based on clustering coefficient, we used the \texttt{edges} and \texttt{triangle} parameters in the statnet package. The \texttt{edges} parameter measures the probability of linkage or no linkage between nodes, and the \texttt{triangle} term looks at the number of triangles or triad formations in the original graph. For the average path length model, \texttt{edges} and \texttt{twopath} terms were used. The  \texttt{twopath} term measures the number of 2-paths in the original network and produces a probability distribution of their formation for the converged ERGM. Lastly, for the assortativity measure, the terms \texttt{edges} and \texttt{degcor} were used to produce the models. The \texttt{degcor} term considers the degree correlation of all pairs of tied nodes (for more on ERGMs see \cite{hunter2008ergm,morris2008specification}). These terms proved to be our best choices for preserving, to a certain extent, the desired structural information. Although the creation of ERGMs is a trial and error process, the selected terms were successful in producing models for each of the original networks.

After a successful model convergence, a simulated graph was generated constraining the number of edges to those of the original graph for each model. It is worth mentioning that within the built-in \texttt{simulate} function in the \texttt{statnet} suite there is no way of forcibly constraining the aspects of the original we want to control. Thus, we experience variation, in some cases more than others. The difference between the original and the simulated graphs seemed more prominent for smaller networks (see Table 1 and Table 2 for comparison) than models based on the larger networks which came closer to the real values of the original graphs.

%.......................................
\subsubsection{Synthetic Labeling}
\label{sec:labeling}
%.......................................

A simple model that parameterizes a labeled graph with a tendency towards homophily (ties disproportionately between those of similar attribute background) is an ``attraction'' model~\cite{Skvoretz2013}.  
In the basic case of a binary attribute variable and a constant tendency to inbreed, two parameters, $p$ and $\tau$, both in the (0,1)  interval, characterize the distribution of ties within and between the two groups.  
The first is the proportion of the population that takes on one value of the attribute (with $1-p$, the proportion taking on the other value).  
The second parameter, the inbreeding coefficient or probability, expresses the degree to which a tie whose source is in one group is "attracted" to a target in that group.  
When $\tau=0$, there is no special attraction and ties within and between groups occur in chance proportions.  
When $\tau>0$, ties occur disproportionately within groups, increasing as $\tau$ approaches 1.  
Given a total number of ties, values for $p$ and $\tau$ determine the number of ties/edges that are between groups, namely, $\delta=|E|\times2\times(1-\tau)p(1-p)$.

In the process of generating synthetic node attributes, we first randomly assign two arbitrary values (i.e., R and B) as labels to all the nodes in the graph for a given $p,1-p$ split.
Then, we draw an R node and a B node at random and swap labels if it would decrease the number of R-B ties.
This process would converge when the total number of cross-group ties reduce to $\delta$ for a particular value of $\tau$.

\begin{figure}[t]
\centering
\includegraphics[scale=0.4]{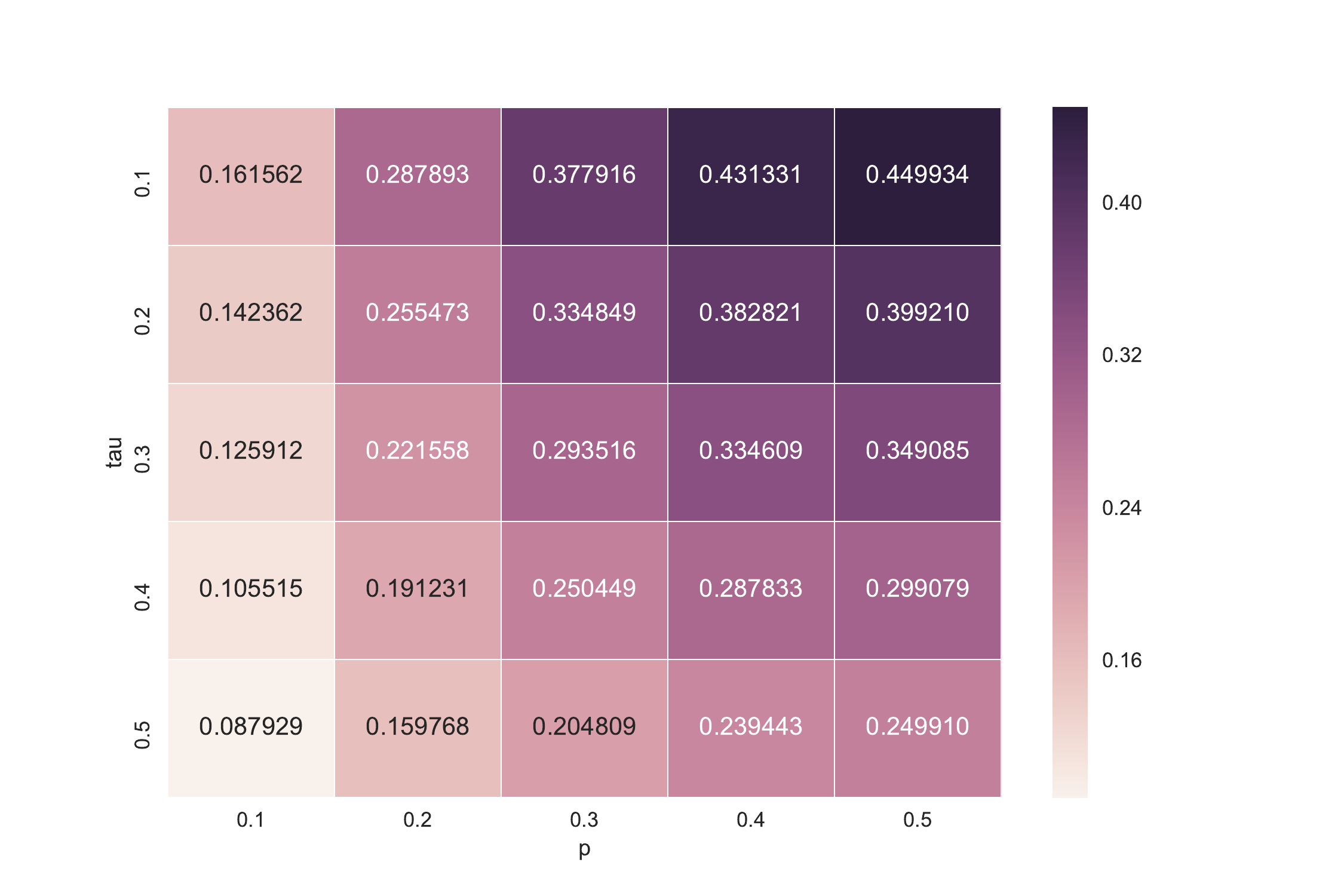}
% 			\label{fig:polblogs_syn_graph}
% % \begin{tabular}{c}
% 	\subfloat[Network: polblogs]{
% 		\includegraphics[width=1\linewidth]{./images/Synthetic-Graphs/polblogs_syn_ties_cross_group.pdf}
% 			\label{fig:polblogs_syn_graph}
% 	}
% 	&
% \subfloat[Network: fb-michigan]{
% \includegraphics[width=0.24\linewidth]{./images/Synthetic-Graphs/fb-michigan_syn_ties_cross_group.pdf}
% 			\label{fig:fb-michigan_syn_graph}
% }
% &
% \subfloat[Network: pokec-1]{
% \includegraphics[width=0.24\linewidth]{./images/Synthetic-Graphs/pokec-1_syn_ties_cross_group.pdf}
% 			\label{fig:pokec-1_syn_graph}
% }
% &
% \subfloat[Network: amazon-products]{
% \includegraphics[width=0.24\linewidth]{./images/Synthetic-Graphs/amazon-products_syn_ties_cross_group.pdf}
% 			\label{fig:amazon-products_syn_graph}
% }
% \\
% \end{tabular}
	 \caption{Network: polblogs, proportion of cross group ties on synthetic networks generated from the original graph structure.
	}
	\label{fig:syn_graph_stat}
\end{figure}

As an example, Figure~\ref{fig:syn_graph_stat} shows the proportion of cross-group ties on the synthetic labelled networks generated from \texttt{polblogs} topology.
The proportion of cross-group ties is proportional to $p$, while it is inversely proportional to $\tau$.
When $p$ reaches its maximum ($p_{max}=0.5$ due to the symmetric nature of binary attribute values), the proportion of cross group ties is relatively larger at minimum inbreeding coefficient $\tau$.  

It should be noted that convergence is not guaranteed for all possible combinations of $p$ and $\tau$.  
The swapping procedure holds constant all graph properties except the mapping of nodes to labels, and consequently, it may not be possible to find a mapping of nodes to labels that achieves a target number of ties between groups (when that number is low as it is for higher values of $\tau$).

Table~\ref{tbl:ERGM-characteristics} presents the graph characteristics of the synthetically generated labeled graphs. 
%.......................................
%\subsubsection{Characteristics of Synthetic Graphs}
%\label{sec:syndatasets}
%.......................................

% \shnote{@Juan, please summarize your results on ERGM graph charactersitics, it's up to you to decide what's the best to show here; E.g., histograms (might be very little information), might be some other way? any plots should be in the same standard with others, use seaborn + python}

% \shnote{Are we talking about convergence issues?}

% with |E| the number of edges in fb.
% we cannot achieve by rearranging labels a graph with a particular value of tau for a given p,1-p split.  First think to note is that convergence becomes a problem as tau increases.  This makes sense because to achieve higher values of tau requires that it be possible to find ever fewer ties between the two groups p, 1-p and to do this by changing labels rather than rewiring.  If we had an erdos-renyi graph, I think we could derive a probability expression that a p,1-p split would have n ore fewer ties between the groups and that would show how hard it might be to achieve a particular tau value.

% \subsection{Training Data}

\begin{table*}[!t]
	\centering
	\caption{Basic statistics of generated ERGM networks, and the population of node pairs. Note that \texttt{dc,cc} and \texttt{apl} define the set of parameters that used to generate ERGM graphs based on assortativity (degree correlation), clustering coefficient, and average path length, respectively. We generated a total of $\approx$ $500$ million identical and non-identical node pairs over three ERGM graph spaces of the six real social network datasets. $S$ is the population of generated node pairs concerning a given graph topology.}
	% 472.6 millions
	\label{tbl:ERGM-characteristics}
	\begin{tabular}{|c|c|c|r|r|r|r|}
		\hline
		Network & ERGM & $d$ & $C$ & $r$ & $\kappa$ & |S| (\textit{millions})\\ \hline
		
		\multirow{3}{*}{polblogs} & dc & 0.02 & 0.03& .08 & 2.52 & 5.5
		\\ \cline{2-7}
		& cc & 0.02 & 0.33& -0.02 & 2.69 & 13.1
		\\ \cline{2-7}
		& apl & 0.02 & 0.10 & -0.06 & 2.49 & 11.5
		\\ 
		\hline
		
			\multirow{3}{*}{fb-caltech} & dc & 0.06 & 0.08 & 0.11 & 2.13 & 1.2
		\\ \cline{2-7}
		& cc & 0.06 & 0.42 & -0.06 & 2.73 & 4.1
		\\ \cline{2-7}
		& apl & 0.06 & 0.07 & 0.11 & 1.97 & 1.2
		\\ 
		\hline
		
		\multirow{3}{*}{fb-dartmouth} & dc & 0.01 & 0.17 & 0.07 & 2.66 & 14.5
		\\ \cline{2-7}
		& cc & 0.01 & 0.24 & 0.04 & 2.77 & 13.2
		\\ \cline{2-7}
		& apl & 0.01 & 0.20 & 0.04 & 2.70 & 14.2
		\\ 
		\hline
		
		\multirow{3}{*}{fb-michigan} & dc & 0.003 & 0.02 & 0.12 & 3.28 & 38.4
		\\ \cline{2-7}
		& cc & 0.002 & 0.20 & 0.12 & 3.52 & 39.9
		\\ \cline{2-7}
		& apl & 0.002 & 0.20& 0.12 & 3.64 & 38.2
		\\ 
		\hline
		
		\multirow{3}{*}{pokec-1} & dc & 2.02E-5 & 0.06 & -0.04 & 5.60 & 29.5
		\\ \cline{2-7}
		& cc & 2.05E-5 & 0.07 & -0.04 & 5.84 & 29.3
		\\ \cline{2-7}
		& apl & 2.04E-5 & 0.06 & -0.04 & 5.63 & 27.3
		\\ 
		\hline
		
		\multirow{3}{*}{amazon-products} & dc & 1.82E-5 & 0.37 & -0.06 & 11.86 & 43.7
		\\ \cline{2-7}
		& cc & 1.82E-5 & 0.40 & -0.06 & 13.52 & 72.5
		\\ \cline{2-7}
		& apl & 1.82E-5 & 0.39 & -0.06 & 13.47 & 74.3
		\\ 
		\hline

		% etc. ...
	\end{tabular}
\end{table*}

\section{Empirical Results}
\label{sec:experiments}

%The re-identification of nodes from naively anonymized graph datasets in which the node identities were obscured has been proven successful over and over again. 
Our objective is not to measure the success of re-identification attacks on original datasets in which node identities have been removed: it has been demonstrated long ago~\cite{backstrom2007wherefore} that naive anonymization of graph datasets does not provide privacy. 
Instead, our objective is to quantify the exposure provided by node attributes on top of the intrinsic vulnerability of the particular graph topology under attack. 

% \ainote{make this consistent with metrics description in section 3}.
% \shnote{Updated.}
%We focus our experimental evaluation on measuring the impact of attribute information to determine the risk of public graph data releases.
%Thus, our objective is to quantify the exposure given by node-attributes on top of the intrinsic vulnerability of the particular graph topology under attack.
%To this end, we are interested in 

In our experiments, we leverage the real and synthetic networks described above.
%For synthetic networks, we first generate ERGM graph topologies with the specified parameters (as described in Section \ref{sec:synthetic_graphs}). Subsequently, we associate synthetic attributes to nodes (as described in Section \ref{sec:synthetic_graphs}) in both original and ERGM graphs. 
We mount the machine learning attack described in Section~\ref{sec:machine-learning-attack} to re-identify nodes using features based on both graph topology and node attributes. 
Our first guiding question is thus: \textit{How much risk of node re-identification is added to a network dataset by its binary node attributes?}

%In Section \ref{sec:vulnerability_cost}, we conduct a statistical analysis to measure the impact of attributes on the task of node re-identification.
%We continue this analysis in Section \ref{sec:diversity_matters} to study the fundamental reasoning behind the vulnerability cost of node attributes.
% with respect to the attraction model we described earlier.
%Finally, we discriminate inherent properties of the network based on attribute and topology level features that facilitate an attacker to re-identify nodes (Section \ref{sec:features}).

%%\subsection{Metrics}

%------------------------------
\subsection{The Vulnerability Cost of Node Attributes}
\label{sec:vulnerability_cost}
%------------------------------

% This includes the discussion of Figure~\ref{fig:org_vulnerability_GS_GS-LBL} in some detail.
% At the end of the discussion: but what makes labeling effect so different? (p and $\tau$ in polblogs vs. others. Compare the fb networks).

\begin{figure*}[htb!]
\scalebox{0.9}{
\begin{tabular}{ccc}
	
	\subfloat[Network: fb-caltech (gender)]{
		\includegraphics[width=0.33\linewidth]{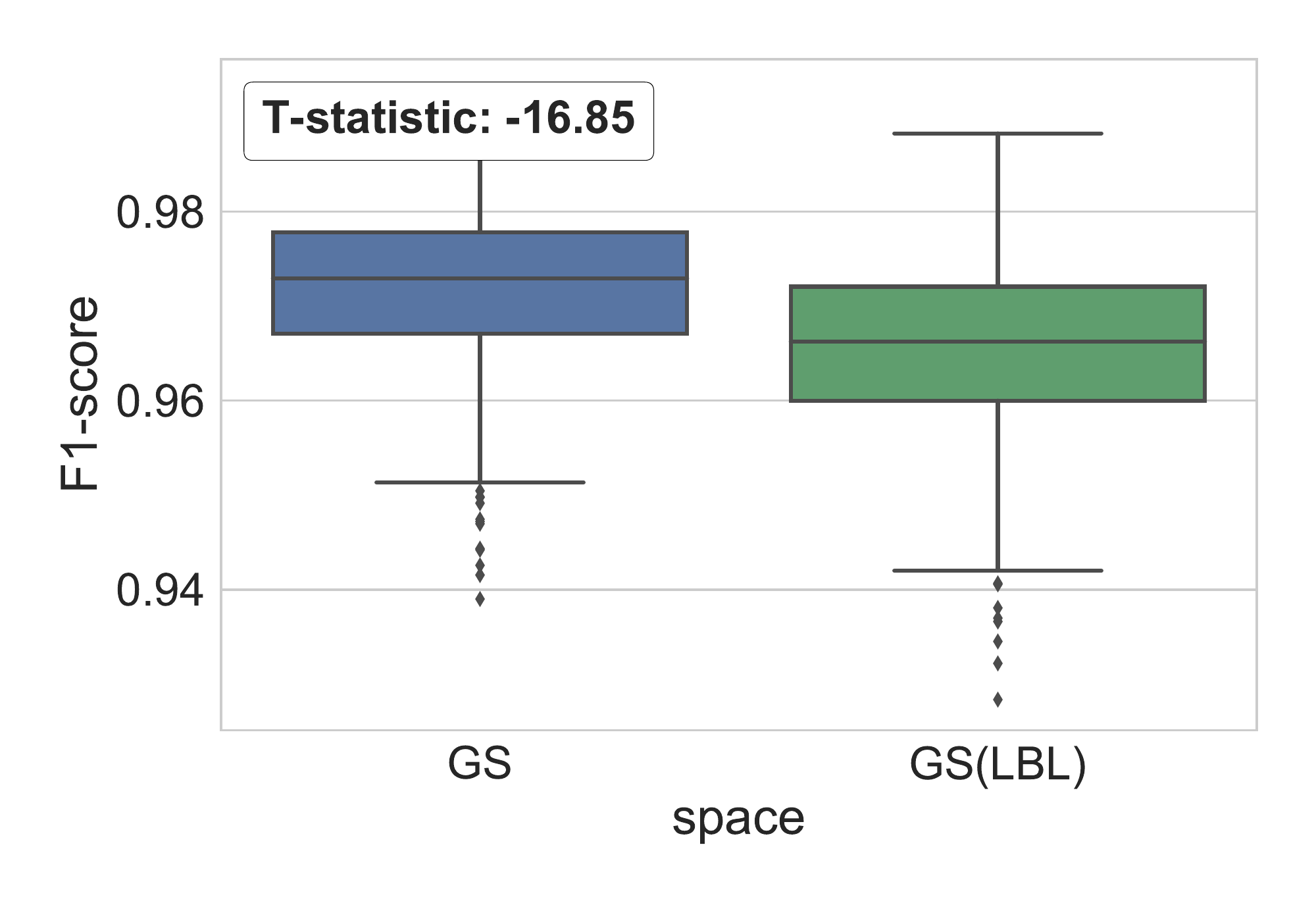}
			\label{fig:fb-caltech-gender_org_vulnerability}
	}
	&
	\subfloat[Network: fb-caltech (occupation)]{
		\includegraphics[width=0.33\linewidth]{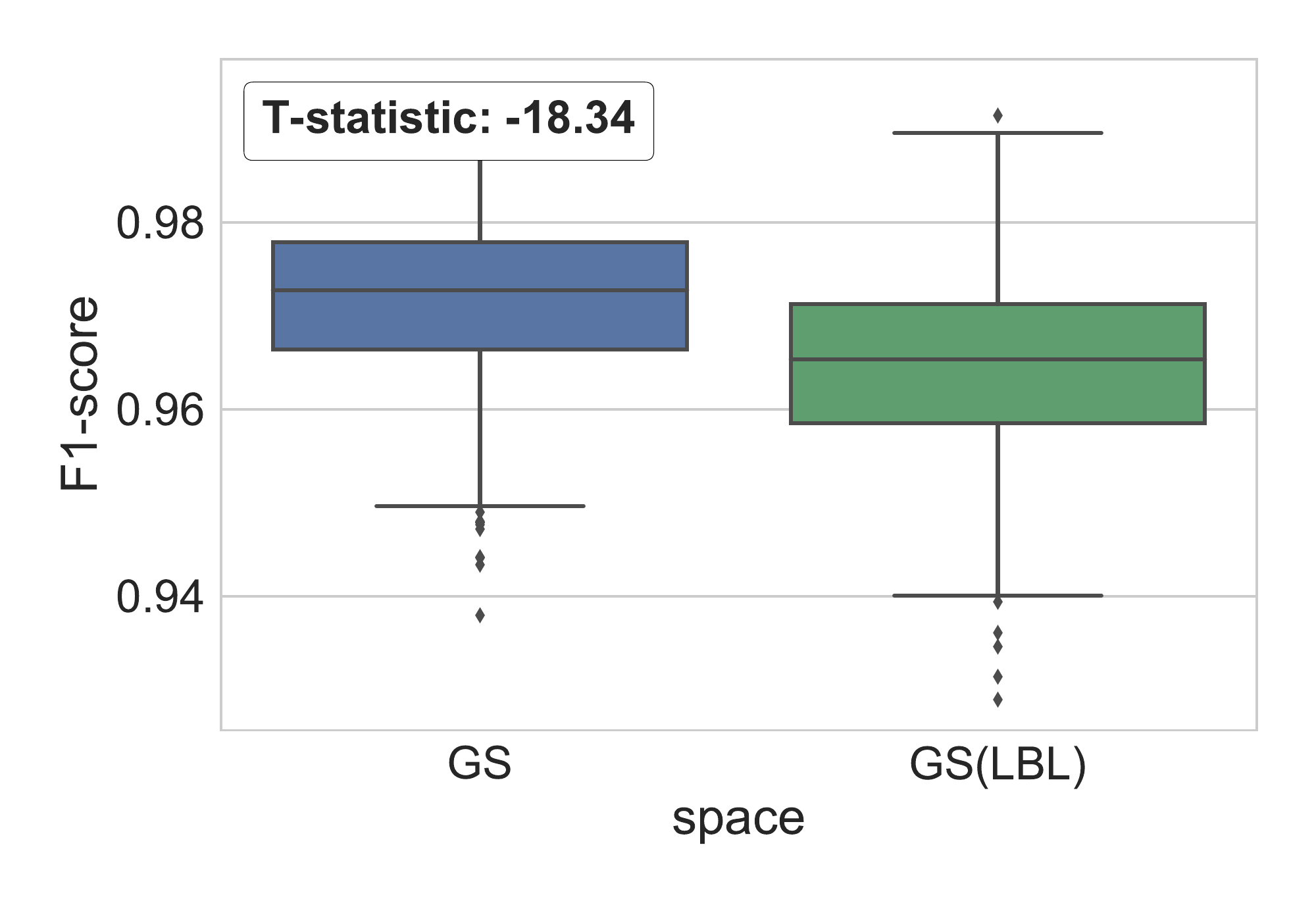}
			\label{fig:fb-caltech-oc_org_vulnerability}
	}
	&\subfloat[Network: polblogs (party)]{
		\includegraphics[width=0.33\linewidth]{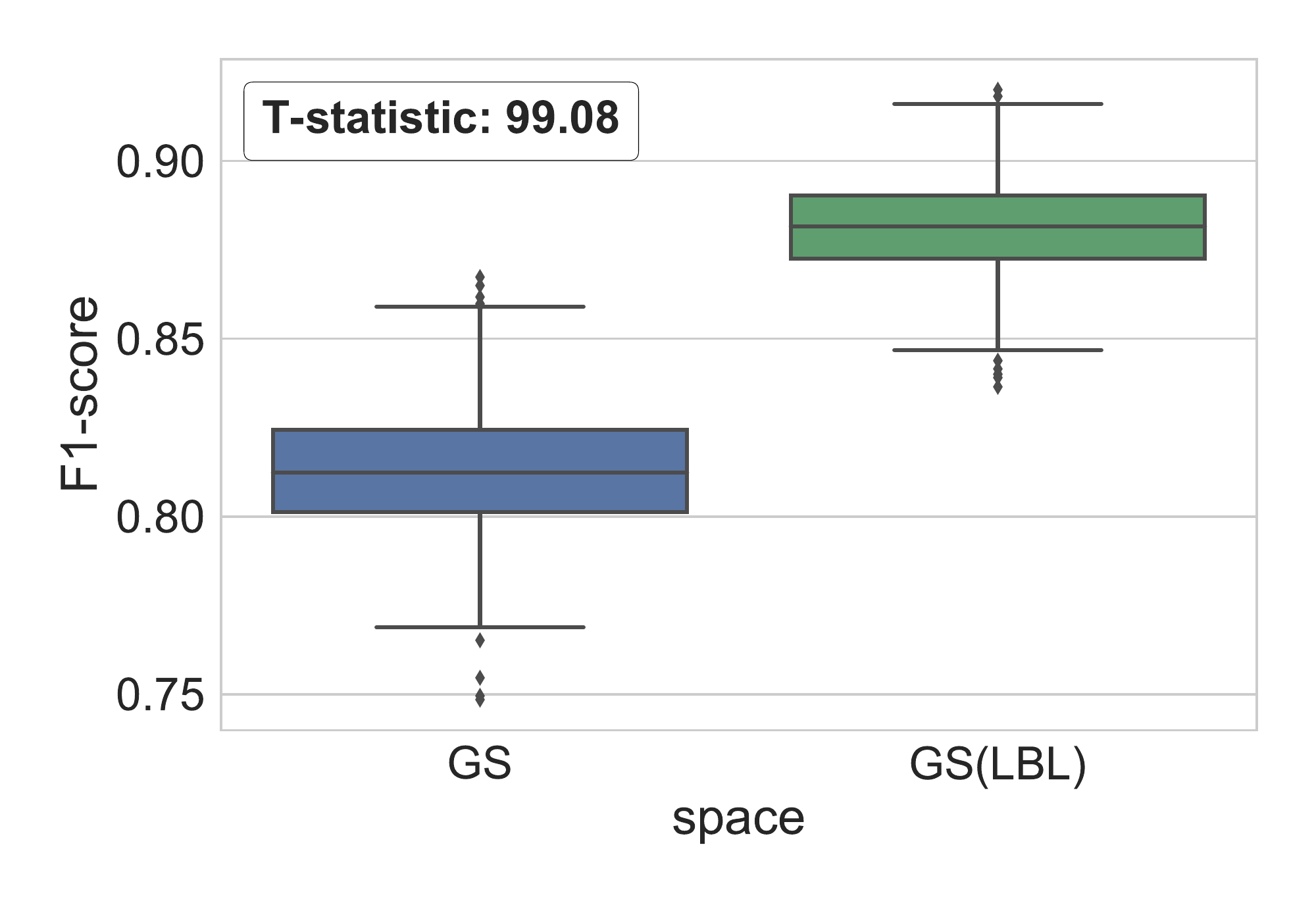}
			\label{fig:polblogs_org_vulnerability}
	}
	\\
	
	\subfloat[Network: fb-dartmouth (gender)]{
		\includegraphics[width=0.33\linewidth]{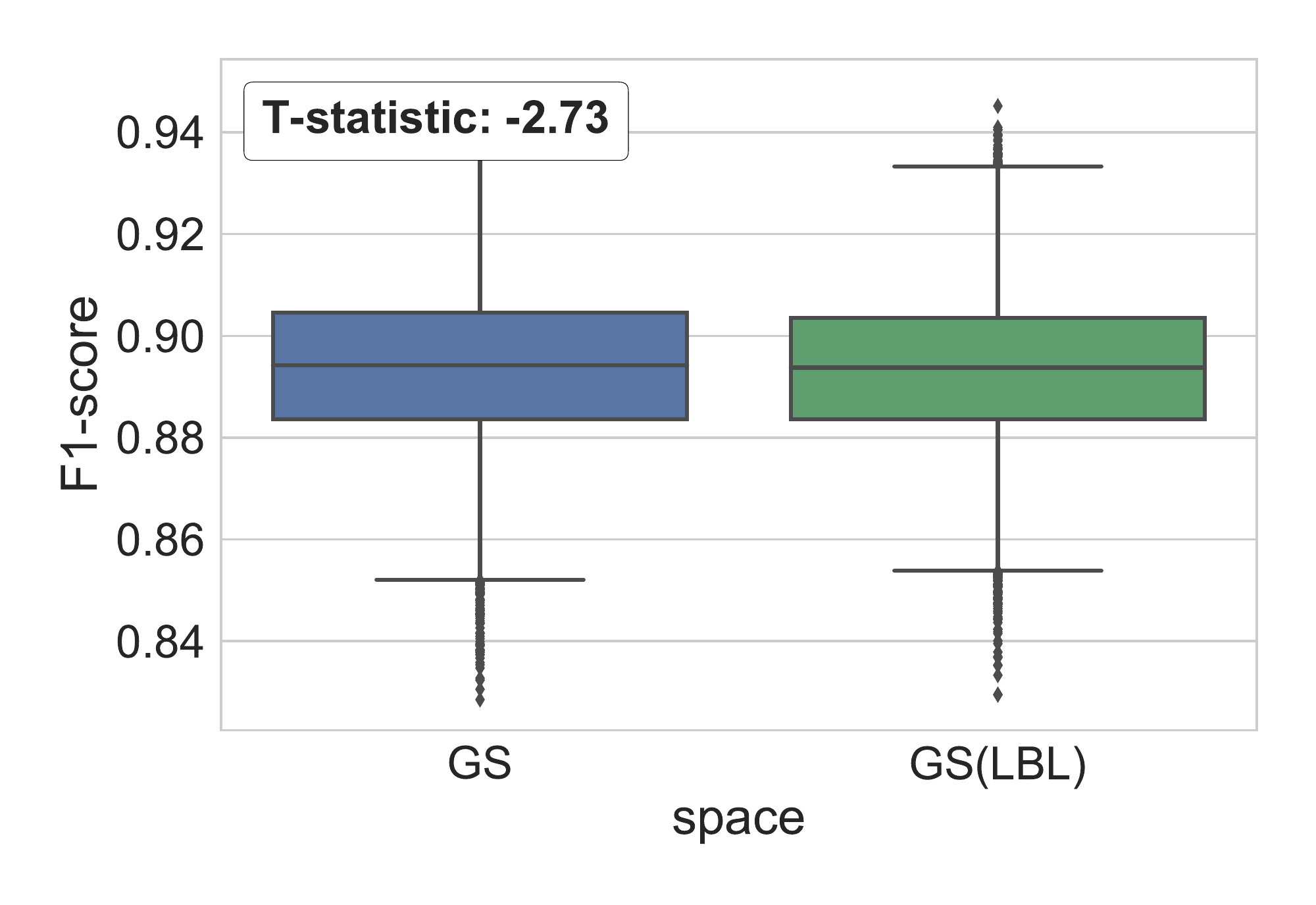}
			\label{fig:fb-dartmouth-gender_org_vulnerability}
	}
	&
	\subfloat[Network: fb-dartmouth (occupation)]{
		\includegraphics[width=0.33\linewidth]{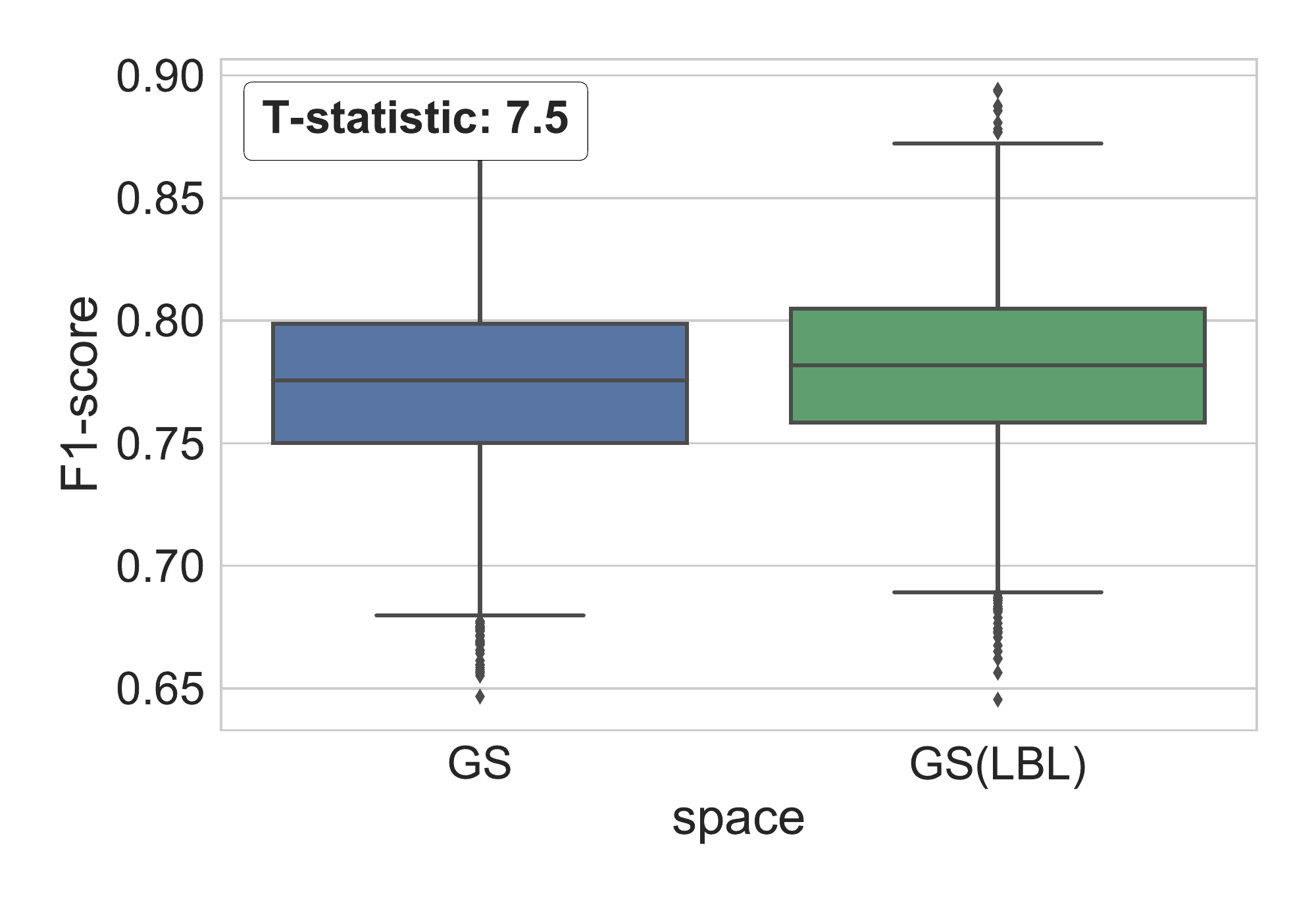}
			\label{fig:fb-dartmouth-oc_org_vulnerability}
	}
	&
	\subfloat[Network: pokec-1 (gender)]{
		\includegraphics[width=0.33\linewidth]{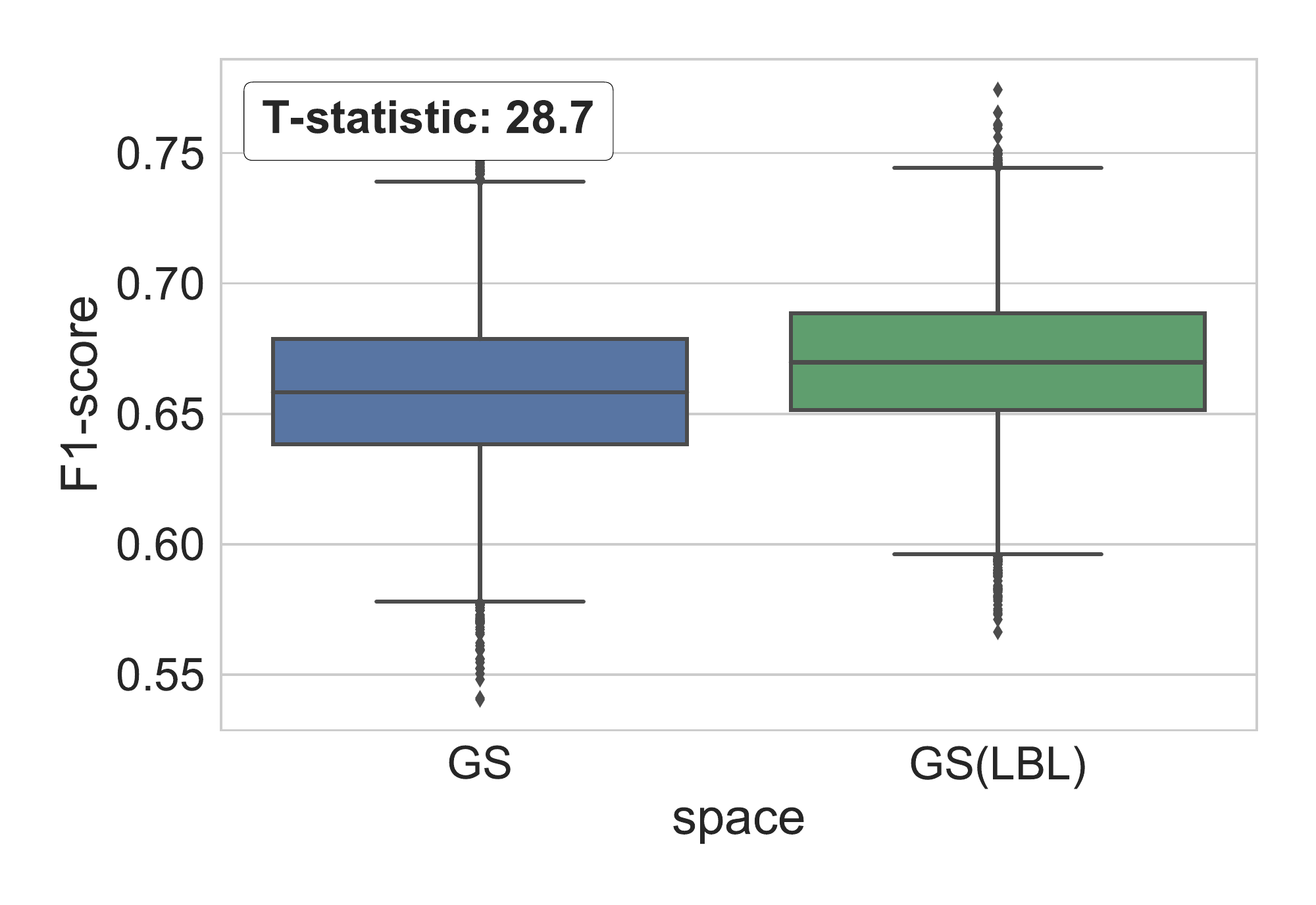}
			\label{fig:pokec-1_org_vulnerability}
	}
	
	\\
	\subfloat[Network: fb-michigan (gender)]{
		\includegraphics[width=0.33\linewidth]{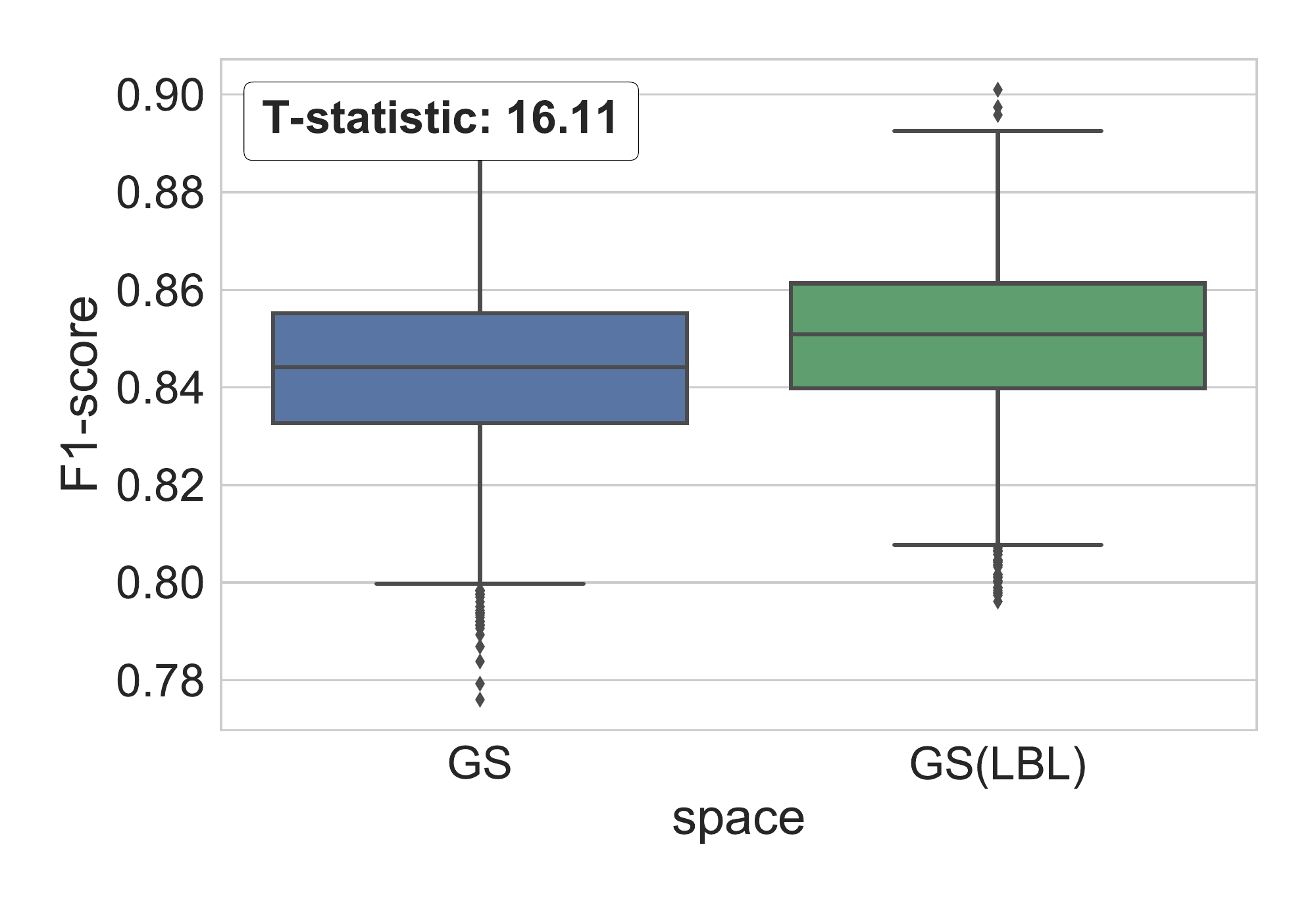}
			\label{fig:fb-michigan-gender_org_vulnerability}
	}
	&
	\subfloat[Network: fb-michigan (occupation)]{
		\includegraphics[width=0.33\linewidth]{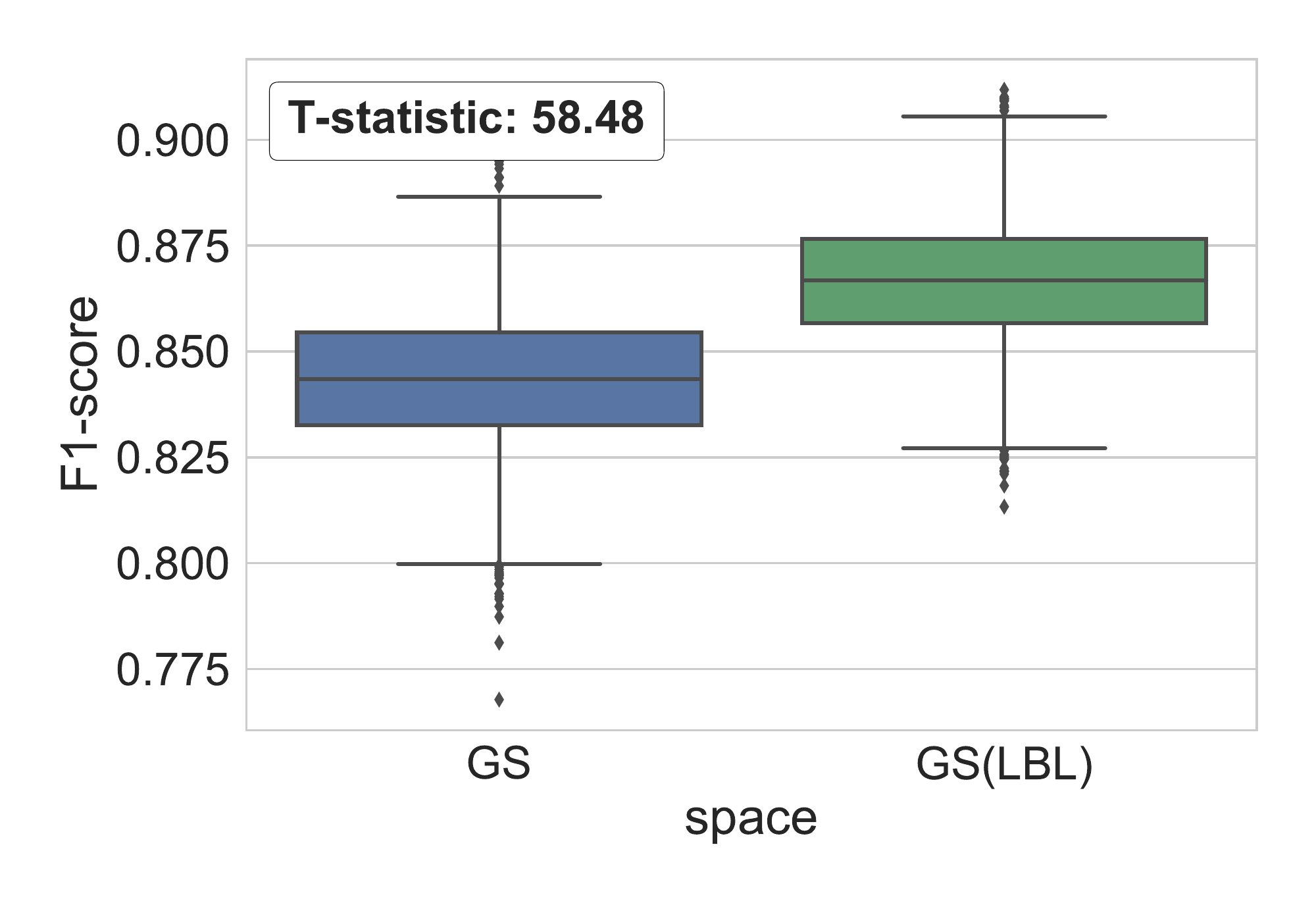}
			\label{fig:fb-michigan-oc_org_vulnerability}
	}
	&
	\subfloat[Network: amazon-products (category)]{
		\includegraphics[width=0.33\linewidth]{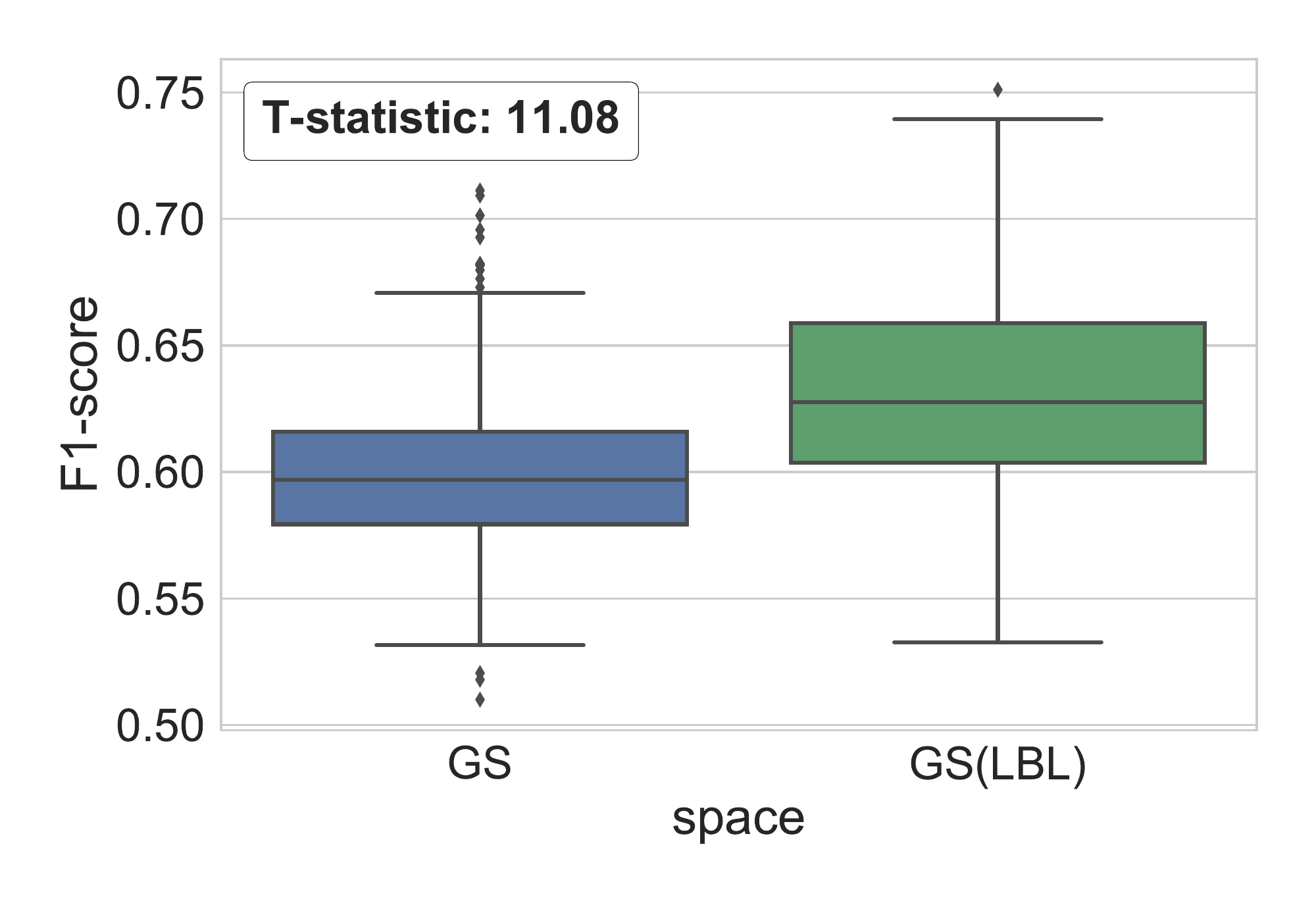}
			\label{fig:amazon-products_org_vulnerability}
	}
	
	\\

\end{tabular}
}
	 \caption{Accuracy of predictions is presented using 5 $\times$ 2 cross-validation F1-scores. Mean accuracy values are shown for real network datasets on GS and GS(LBL) along with the T-statistic which describes the difference in means of the GS and GS(LBL) vectors of prediction probabilities statistically. The network with node attributes is more vulnerable to node re-identification when T-statistic is positive and large.
% 	 on the topological parameters of \texttt{dc} (degree-correlation), \texttt{cc}(clustering coefficient) and \texttt{apl} (average path length).
	}
	\label{fig:org_vulnerability_GS_GS-LBL}

\end{figure*}

% \flush

Figure~\ref{fig:org_vulnerability_GS_GS-LBL} presents the accuracy of node re-identification in the original graph topology GS and in the same topology augmented with node attributes GS(LBL). 
As expected, the re-identification attack performs (generally) better when node attributes are used in the attack. 
Surprising to us, however, is the relatively small vulnerability cost that node attributes introduce. 
For example, the \emph{occupation} attribute has a barely noticeable benefit to the attacker in \texttt{fb-dartmouth}. 
More interestingly, however, the same attribute performs differently for the other two Facebook networks considered: for \texttt{fb-caltech} the \emph{occupation} label functions as noise, leading to a small decrease in the F1-score. 
For \texttt{fb-michigan}, on the other hand, the \emph{occupation} label significantly improves the attacker's performance. 

Another observation from this figure is that different node attributes applied to the same topology have different outcomes: see, for example, the case of the \texttt{fb-michigan} topology, where the difference between the impacts of the \emph{gender} and the \emph{occupation} attributes is the largest. 
We thus formulate a new question: \emph{What placement of attributes onto nodes reveal more information?}

%Note that, while the Y-axes in plots are different for visibility, the added vulnerability is best seen from the T-statistics values reported in each case.

%------------------------------
\subsection{Diversity Matters, Homophily Not}
\label{sec:diversity_matters}
%------------------------------

To understand how the placement of attribute values on nodes affects vulnerability, we generate synthetic node attributes in a controlled manner. 
By varying $p$ (the diversity ratio) and $\tau$ (the bias of nodes with same-value attributes to be connected by an edge), we can study the effect of these parameters on node re-identification. 

\begin{figure*}[!]
	\centering
	\subfloat[Network: polblogs
% 	, original network: T-statistic (gender): $99.08$
]{
		\includegraphics[scale=0.3]{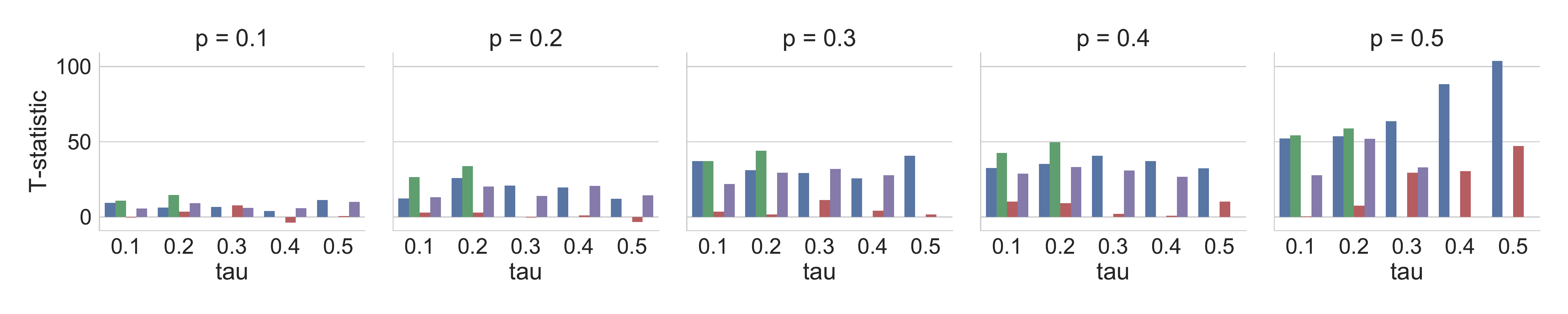}
			\label{fig:polblogs_syn_vulnerability_TStat_Barplot}
	}
	\hspace{0mm}
% \subfloat[Network: fb-caltech
% % original network: T-statistic (gender): $-16.85$, T-statistic (occupation): $-18.34$
% ]{
% 	\includegraphics[scale=0.3]{./images/T-stat/fb-caltech_syn_vulnerability_TStat_Barplot_updated.pdf}
% 			\label{fig:fb-caltech_syn_vulnerability_TStat_Barplot}
% }
% \hspace{0mm}	
	
\subfloat[Network: fb-dartmouth
% , original network: T-statistic (gender): $-2.73$, T-statistic (occupation): $7.5$
]{
\includegraphics[scale=0.3]{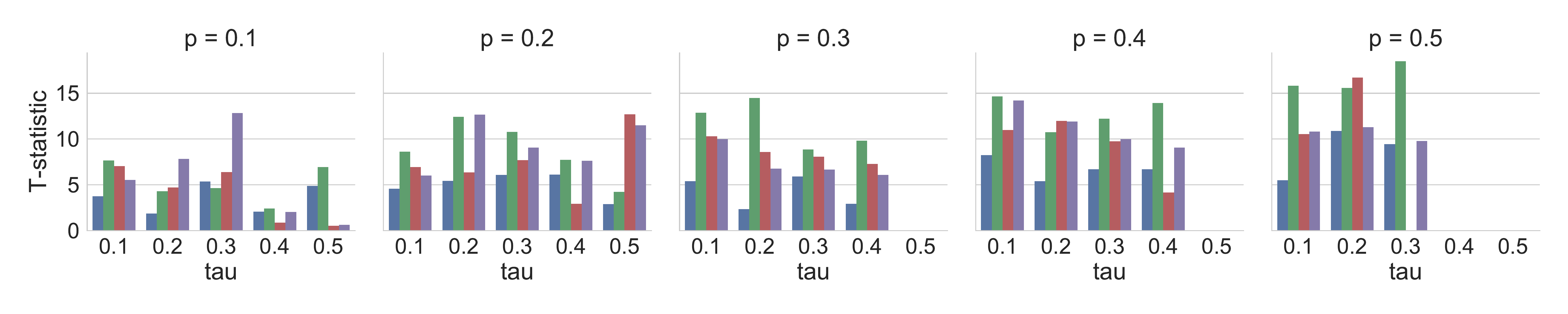}
			\label{fig:fb-dartmouth_syn_vulnerability_TStat_Barplot}
}

\hspace{0mm}	
	
\subfloat[Network: fb-michigan
% , original network: T-statistic (gender): $16.11$, T-statistic (occupation): $58.48$
]{
\includegraphics[scale=0.3]{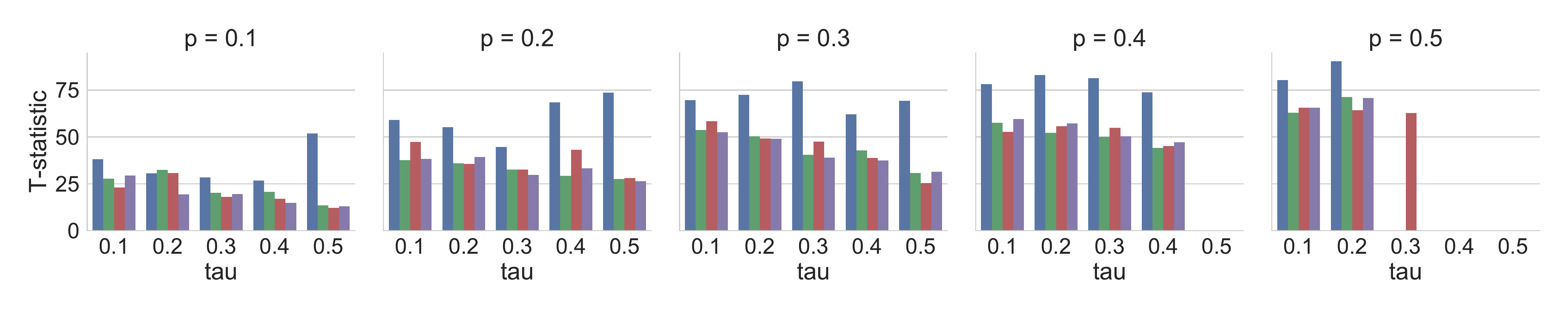}
			\label{fig:fb-michigan_syn_vulnerability_TStat_Barplot}
}
\hspace{0mm}

\subfloat[Network: pokec-1
% , original network: T-statistic (gender): $16.11$, T-statistic (occupation): $58.48$
]{
\includegraphics[scale=0.3]{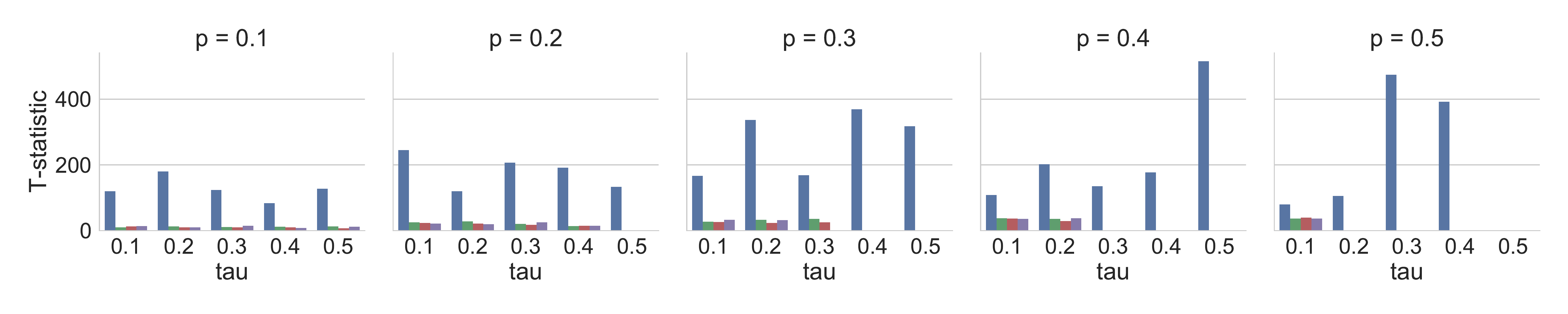}
			\label{fig:pokec-1_syn_vulnerability_TStat_Barplot}
}
\hspace{0mm}

\subfloat[Network: amazon-products
% , original network: T-statistic (category): $11.08$
]{
\includegraphics[scale=0.3]{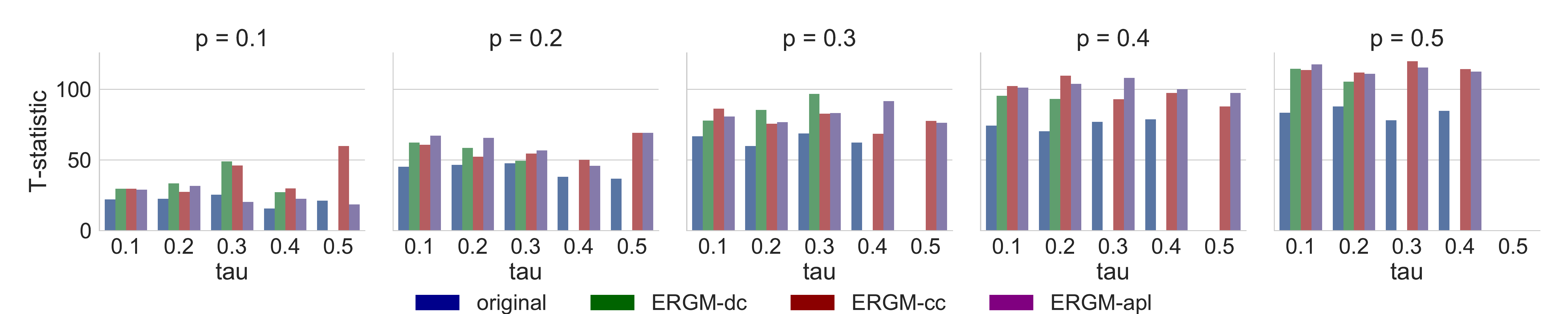}
			\label{fig:amazon-products_syn_vulnerability_TStat_Barplot}
}
	 \caption{Accuracy of predictions is presented using 5 $\times$ 2 cross-validation F1-scores. T-statistic between prediction scores of GS(LBL) and GS are shown. 
	 % 	 GS represents the accuracy of predictions from the features based on the topology and GS (LBL) represents the accuracy of predictions from the features based on both the topology and node attributes. 
Results are shown across different structures of original and ERGM graphs. 
Each ERGM graph is presented using the generated parameters of \texttt{dc} (degree-correlation), \texttt{cc}(clustering coefficient) and \texttt{apl} (average path length)
	}
	\label{fig:syn_vulnerability_TStat}
\end{figure*}

Figure~\ref{fig:syn_vulnerability_TStat} presents the T-statistics of the F1-scores for node re-identification attacks on the original topology vs. labeled versions of the original topology. 
In addition to the original topologies, Figure~\ref{fig:syn_vulnerability_TStat} also presents results on various synthetic networks generated as presented in Section~\ref{sec:synthetic_graphs}. % new topologies are generated using ERGMs to isolate particular graph characteristics of the original datasets, and then labeled according to the same process as for the original topology. 

We observe three phenomena: First, it appears that $p$ is positively correlated with the T-statistic value measuring the re-identification impact of attributes.
That is, the more diversity (that is, the larger $p$), the more vulnerable to re-identification the labeled nodes become on average. 
Intuitively, in a highly skewed attribute population, while the minority nodes will be identified quicker due to node attributes, the majority remains protected. 
On the other hand, when $p=0.5$, a network has two equal-sized sets of nodes where each set takes one of two attribute values.
This is explained by the fact that the NAD feature vector captures more diverse information in the attributes of neighbots when $p$ is larger. 
This is also the explanation for why the node attributes contribute so much more to vulnerability in the \texttt{polblogs} dataset, which has a large diversity ($p=0.48$) (thus, almost equal numbers of conservative and liberal blogs).
Note that the effect of $p$ on the added vulnerability remains consistent across all topologies (real and synthetic) tested.

The second observation is that there is no visible pattern on how $\tau$  influences the vulnerability added by binary node attributes. 
While this is disappointing from the perspective of story telling, it is potentially encouraging for data sharing, as it suggests that datasets that record homophily (or influence, the debate is irrelevant in this context) do not have to be anonymized by damaging this pattern.
As a specific example, the privacy of a dataset that records an information dissemination phenomenon could be provided without perturbing the cascading-related ties. 

The third class of observations is related to the relative effect of the topological characteristics on the added vulnerability. 
Both \texttt{amazon-products} and \texttt{pokec-1} are orders of magnitude sparser than the other datasets considered. 
This means that the topological information available to the machine learning algorithm is limited. 
In this situation, the addition of the attribute information turns out to be very significant: the T-statistic values for these datasets are significantly larger than for the other datasets, with values over 400 in some cases.

Another topological effect is noticed when comparing the real \texttt{pokec-1} topology with the ERGM-generated ones in Figure~\ref{fig:pokec-1_syn_vulnerability_TStat_Barplot}: the node attribute contributes much more to the vulnerability of the original topology compared to the synthetic topologies. 
The reason for this unusual behavior may lay in the different clustering coefficients of the networks, as seen in Tables~\ref{tbl:dataset} and \ref{tbl:ERGM-characteristics}: the ERGM-generated topologies have clustering coefficients one order of magnitude higher than the original topology (for the same graph density), which leads to more diverse NDD feature vectors for the networks with higher clustering and thus richer training information. 
This in turn leads to better accuracy in node re-identification in the unlabeled ERGM topologies (with higher clustering) than in the original topology. 
For example, the maximum F1-score for the ERGM-dc topology is 0.92 while for the original is 0.76 in \texttt{pokec-1}. 
Thus, the relative benefit of the node attribute is significantly higher when the topology features were poorer.

%The relative rank (i.e., depth) of a feature used as a decision node in a tree can be used to assess the relative importance of that feature with respect to the predictability of the target variable. Features used at the top of the tree contribute to the final prediction decision of a larger fraction of the input samples. The expected fraction of the samples they contribute to can thus be used as an estimate of the relative importance of the features.

%------------------------------
\subsection{Topology Leaks}
\label{sec:features}
%------------------------------

\begin{figure*}[!ht]
\scalebox{0.9}{
\begin{tabular}{ccc}

	\subfloat[Network: fb-caltech (gender)]{
		\includegraphics[width=0.33\linewidth]{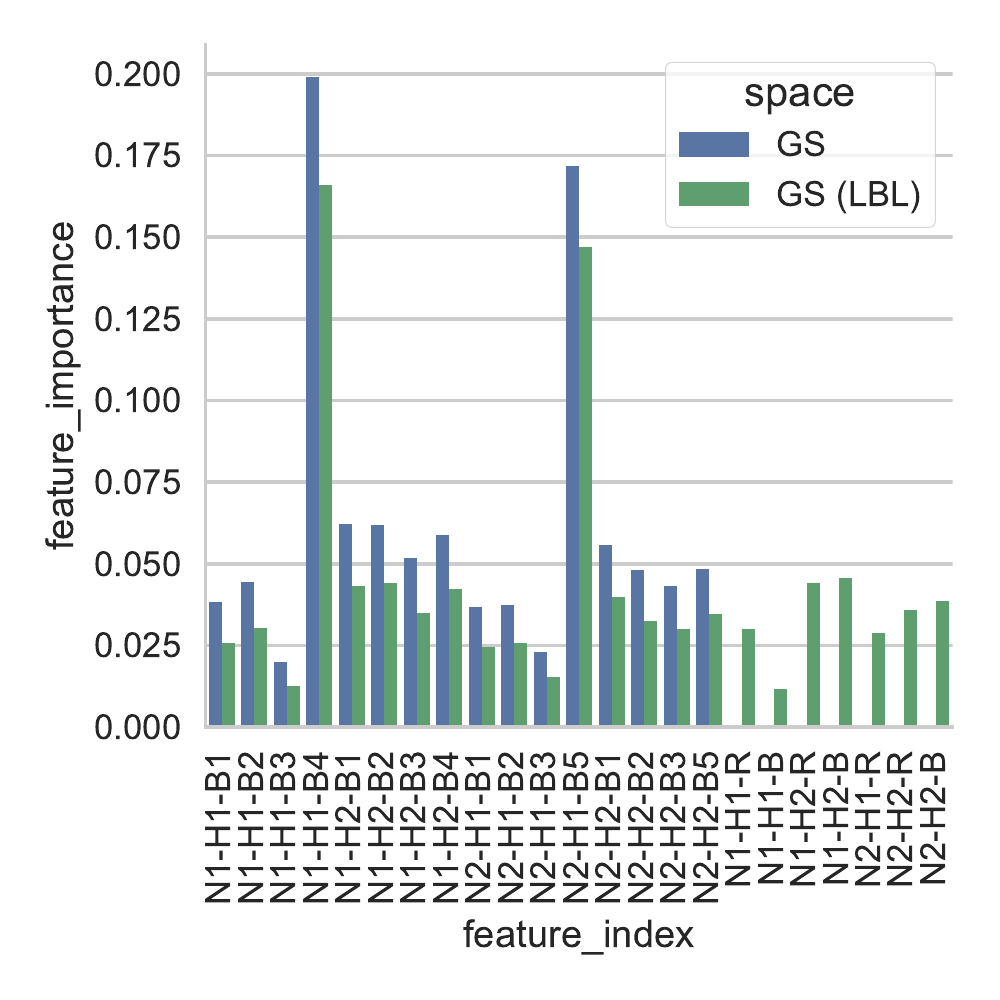}
			\label{fig:fb-caltech_org_vulnerability_features_gender}
	}
&

	\subfloat[Network: fb-caltech (occupation)]{
		\includegraphics[width=0.33\linewidth]{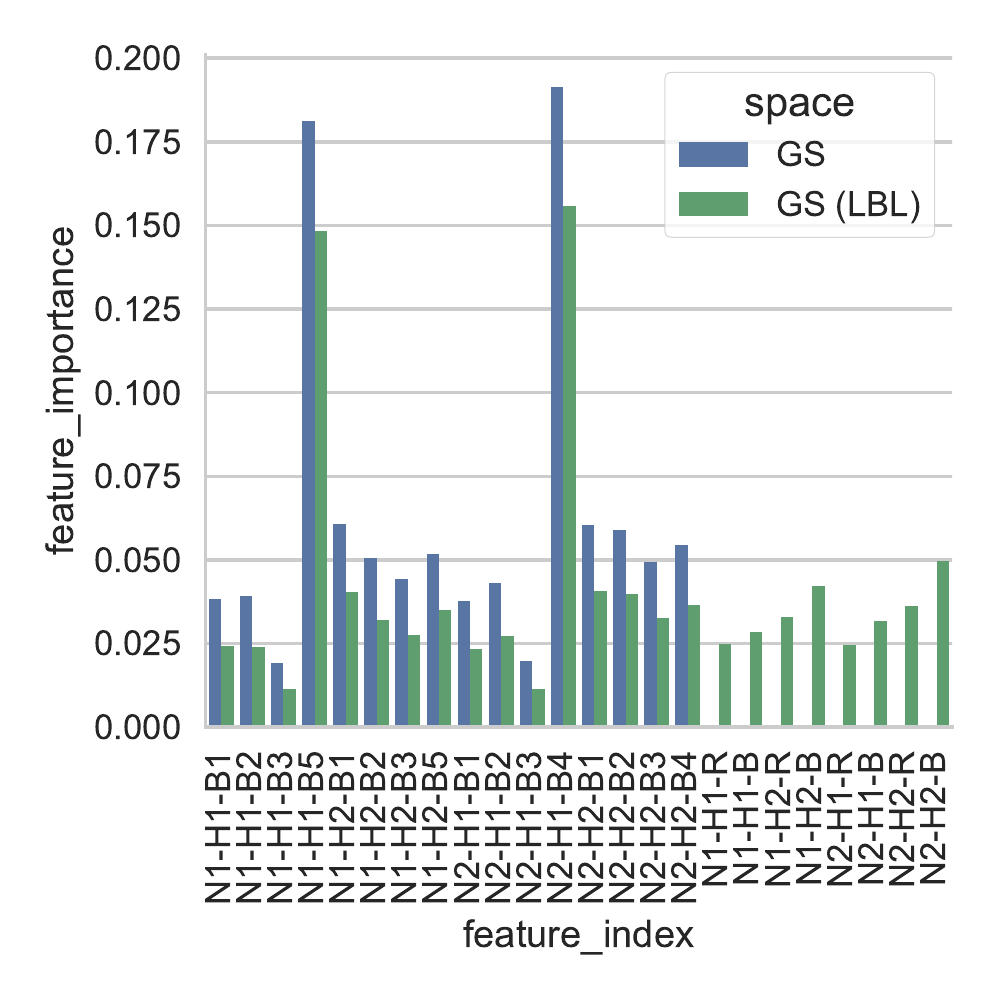}
			\label{fig:fb-caltech-oc_org_vulnerability_features_occupation}
	}
	&
	\subfloat[Network: polblogs (party)]{
		\includegraphics[width=0.33\linewidth]{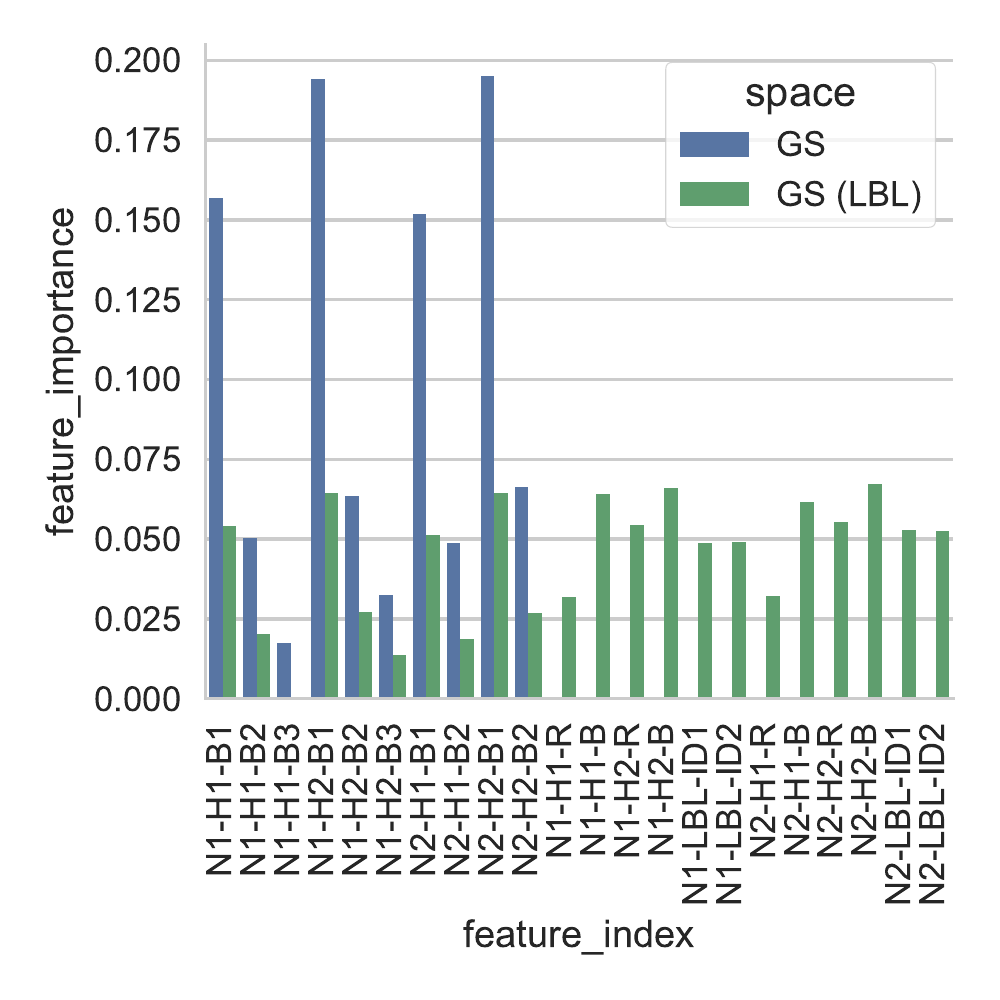}
			\label{fig:polblogs_org_vulnerability_features}
	}
\\

	\subfloat[Network: fb-dartmouth (gender)]{
		\includegraphics[width=0.33\linewidth]{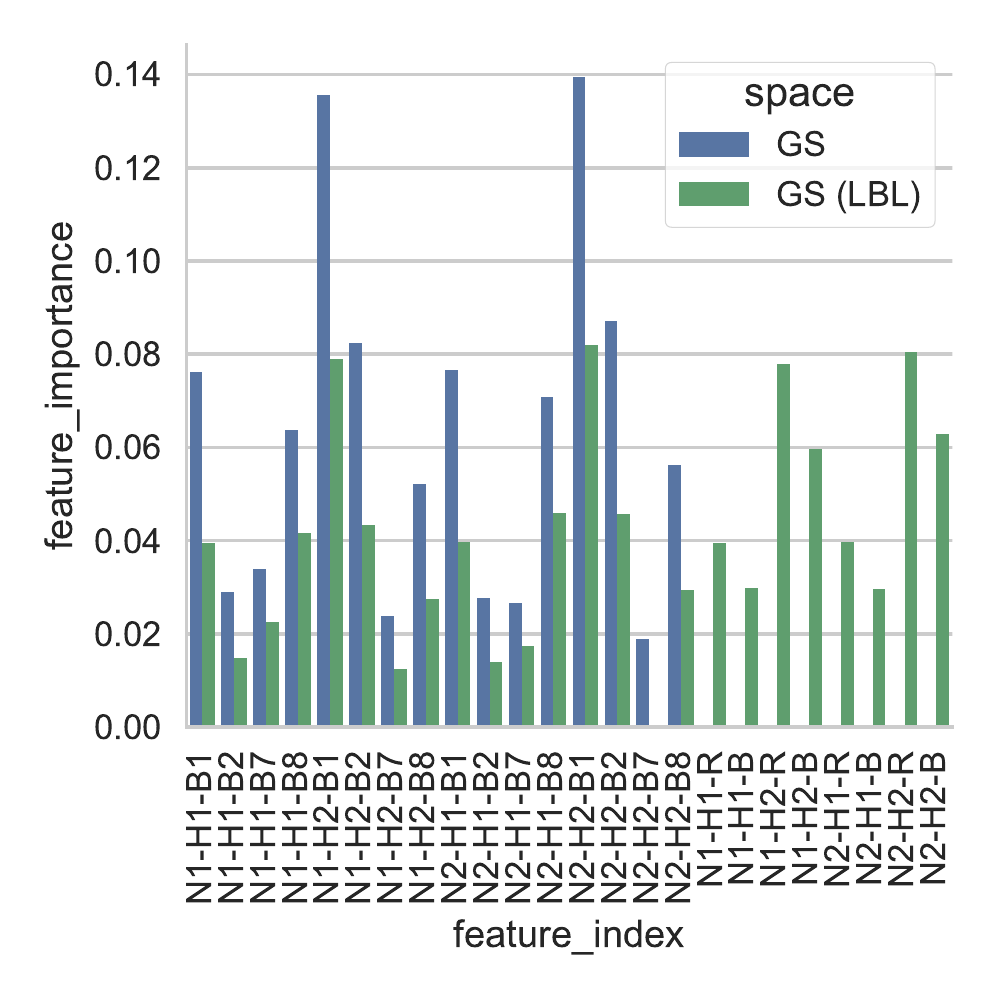}
			\label{fig:fb-dartmouth_org_vulnerability_features_gender}
	}
&
	
	\subfloat[Network: fb-dartmouth (occupation)]{
		\includegraphics[width=0.33\linewidth,height=0.25\textheight, keepaspectratio]{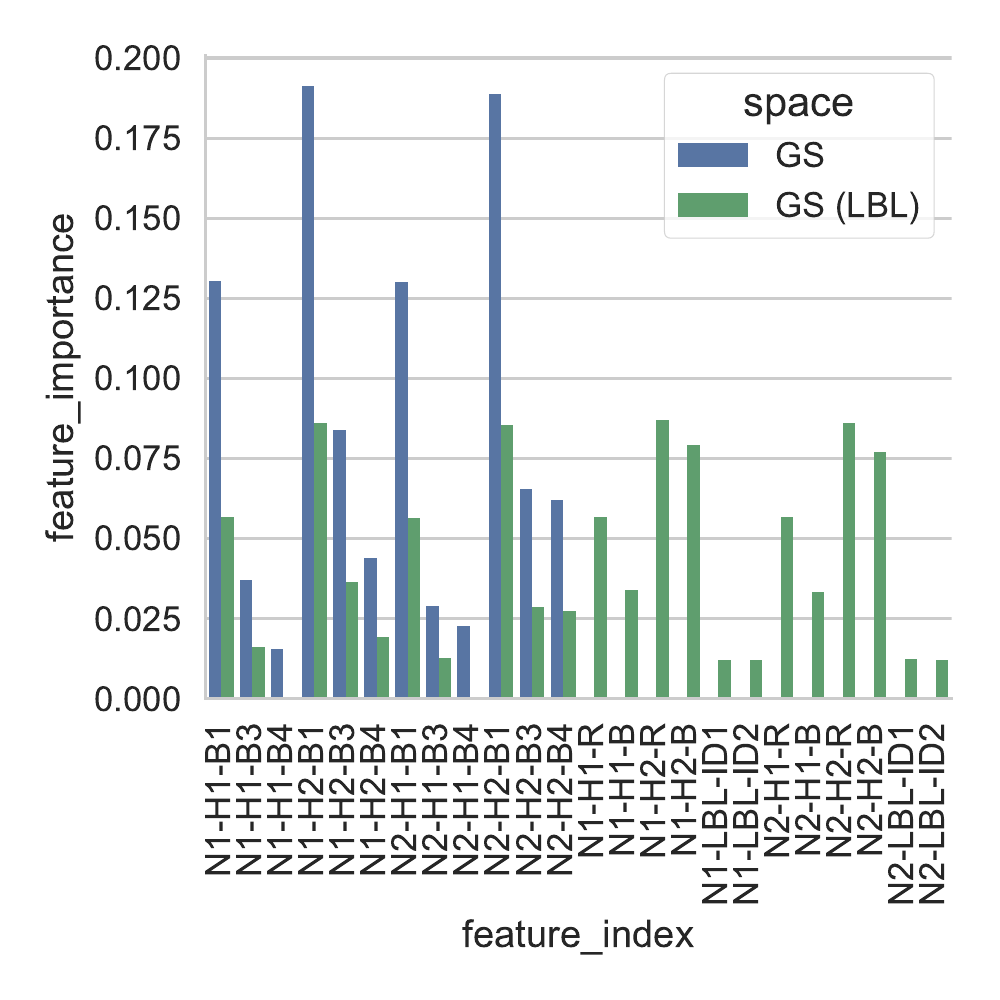}
			\label{fig:fb-dartmouth-oc_org_vulnerability_features}
	}
&
\subfloat[Network: pokec-1 (gender)]{
		\includegraphics[width=0.33\linewidth,height=0.25\textheight, keepaspectratio]{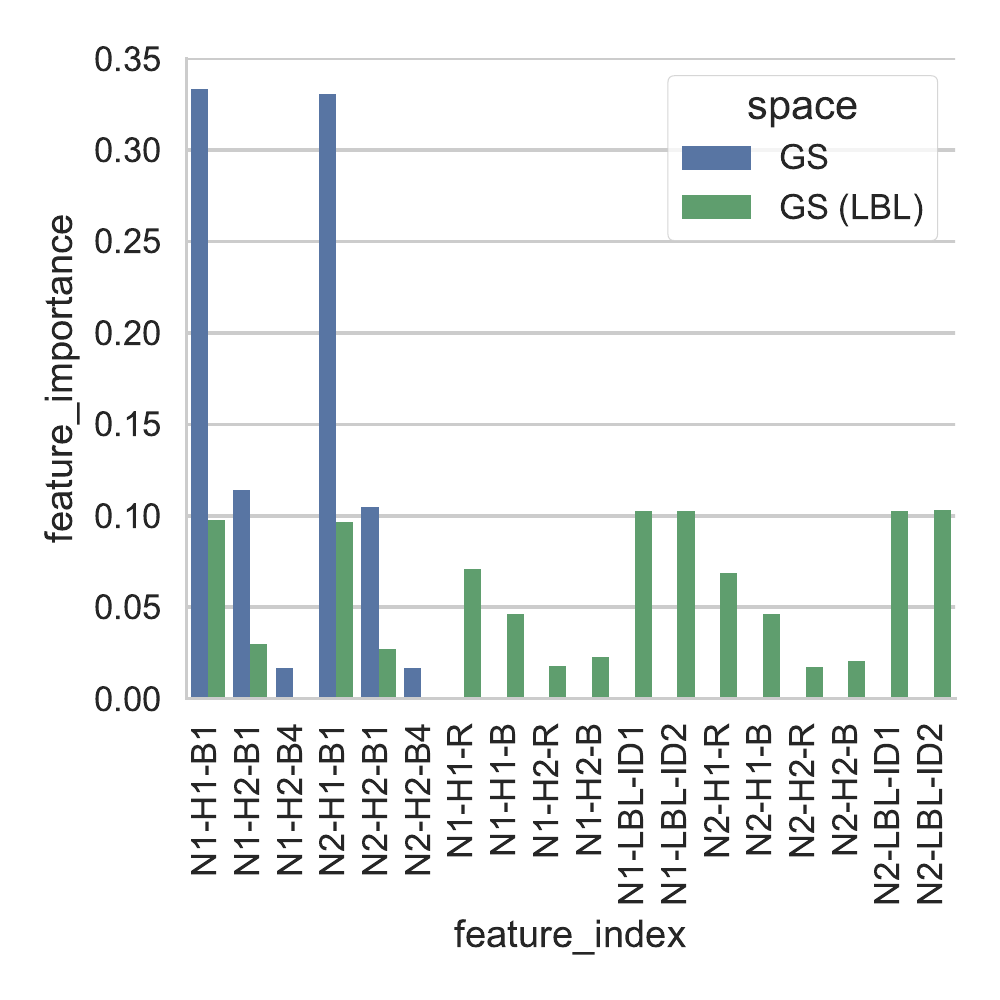}
			\label{fig:pokec-1_org_vulnerability_features}
	}
	\\

	\subfloat[Network: fb-michigan (gender)]{
		\includegraphics[width=0.33\linewidth,height=0.25\textheight, keepaspectratio]{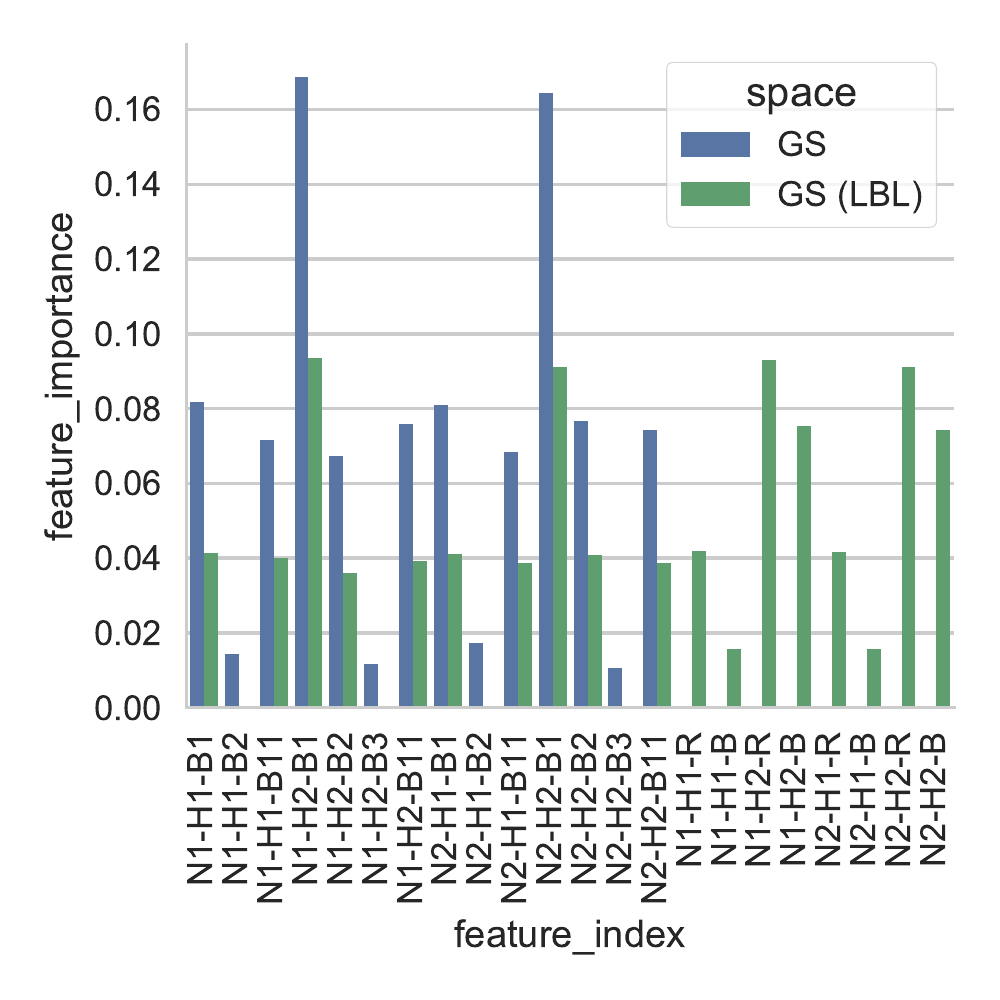}
			\label{fig:fb-michigan-gender_org_vulnerability_features}
	}
&
	\subfloat[Network: fb-michigan (occupation)]{
		\includegraphics[width=0.33\linewidth,height=0.25\textheight, keepaspectratio]{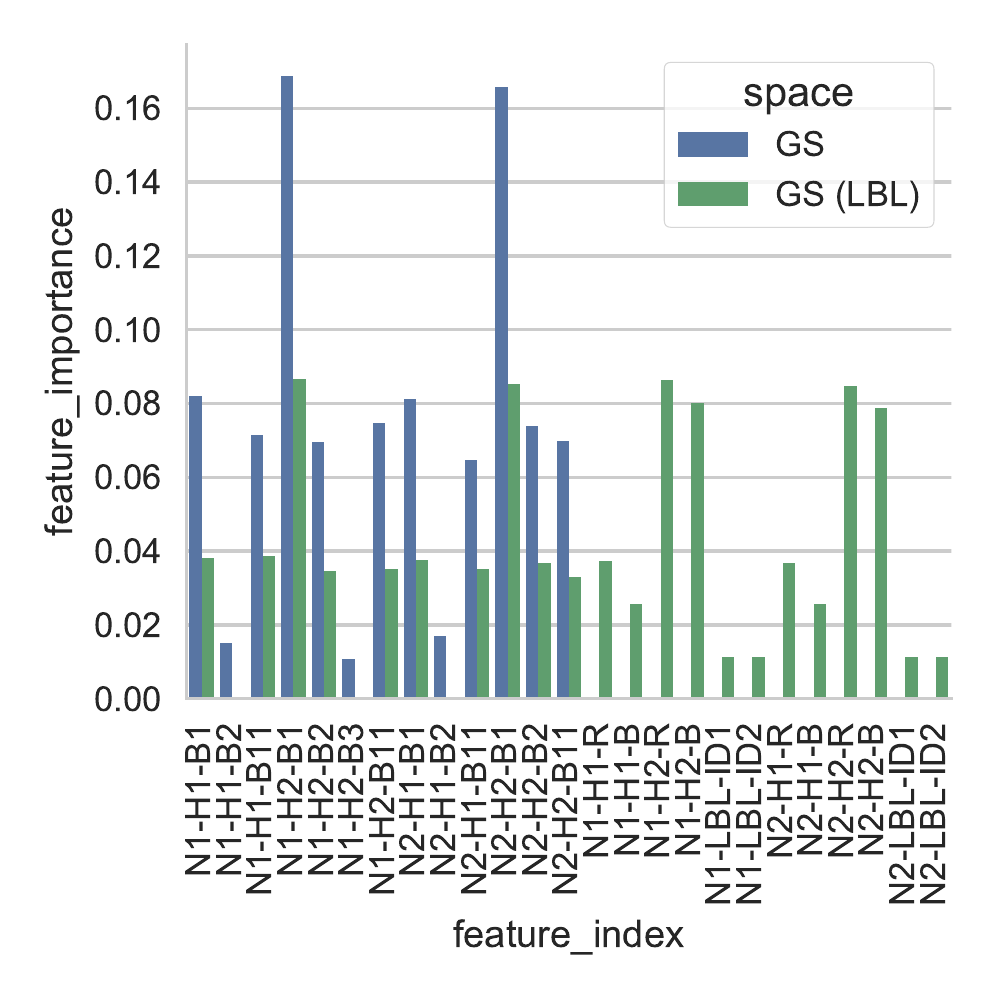}
			\label{fig:fb-michigan-oc_org_vulnerability_features}
	}
&

	\subfloat[Network: amazon-products (category)]{
		\includegraphics[width=0.33\linewidth,height=0.25\textheight, keepaspectratio]{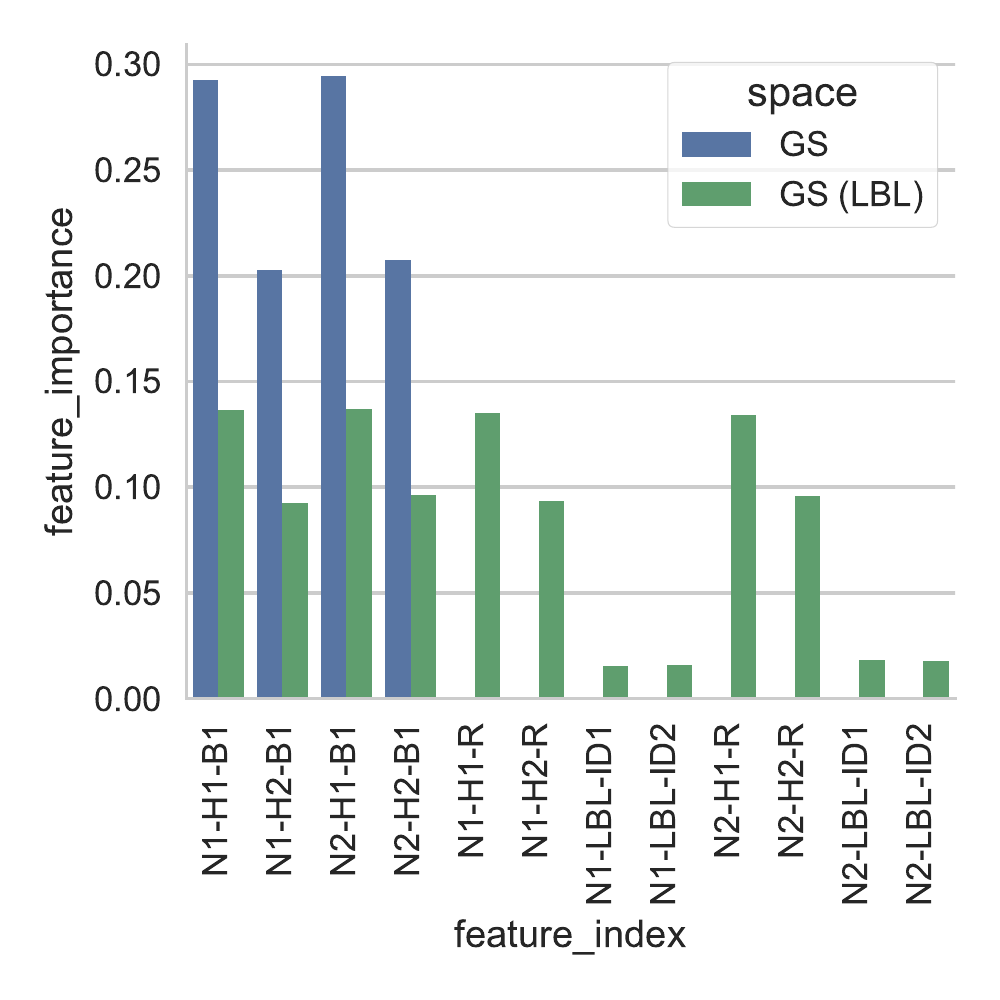}
			\label{fig:amazon-products_org_vulnerability_features}
	}
	\\
\end{tabular}
}
	 \caption{Probability distribution of the feature importance scores across original networks. NDD features are presented in the index order of node (N), hop (H) and bin (B). As an example, the feature \texttt{N1-H2-B1} presents the first bin of the $NDD^2_1[k]$ vector. NAD features are presented in the index order of node (N), hop (H) and binary attribute value $\in {R,B}$. As an example, the feature \texttt{N1-H2-R} presents $NAD^2_1[R]$. Any feature that does not contribute to the final prediction decision with at least $1\%$ of the samples in average is omitted. 
% 	 GS denotes the features based on the topology and GS (LBL) denotes the features based on both the topology and node attributes.
	}
	\label{fig:vulnerability_features_org}
\end{figure*}

Figure~\ref{fig:vulnerability_features_org} presents the importance of features that are used in node re-identification. 
A high importance score represents a feature that is responsible for accurately classifying a large proportion of examples.

We make three observations from this figure. 
First, most of the NAD features (together with node's attribute value) that represent node attribute information prove to be important in all datasets. 

Second, among the NDD features, only a small number contributes consistently to accurate prediction. 
As shown in Figures~\ref{fig:polblogs_org_vulnerability_features}~-~\ref{fig:amazon-products_org_vulnerability_features}, the first bin of 1-hop and 2-hop NDD vectors contribute the most.
That is, a high impact on the re-identification of a node is brought by the number of its neighbors with degrees between 1 and 50. 
Even in large networks such as \texttt{pokec-1} and \texttt{amazon-products} with a larger range of node degrees, this behavior is observed. 

Third, Figure~\ref{fig:vulnerability_features_org} suggests what features explain the effect of diversity $p$ on node re-identification in labeled networks. 
On datasets with large diversity (such as \texttt{polblogs} or \texttt{pokec-1}), the topological information contributes less than on datasets with low diversity (such as \texttt{fb-caltech (gender)}).
This is because high diversity correlates to richer NAD feature vectors, and thus the relative importance of the NAD features increases.

% when p is high, less topological contribution

% \clearpage

\section{Summary and Discussions}
\label{sec:discussions}

Our work shows that the addition of even a single binary attribute to nodes in a network graph increases its vulnerability to re-identification.  Previous work showed that vulnerability increases with the addition of multiple, multi-category attributes~\cite{JiMittal2016De-SAG}. 
We measure the vulnerability increase and study how it is affected by network and attribute properties. 

The increase in vulnerability derives from the fact that the machine learning attack makes use of the interaction between topology and the distribution of node labels. Using information about the distribution of labels in a node's neighborhood provides additional leverage for the re-identification process even when labels are rudimentary.

Furthermore, we find that a population's diversity on the binary attribute consistently degrades anonymity and increases vulnerability.  Diversity means a more even distribution of the binary attribute which produces a more varied set of neighborhood distributions that a particular node may exhibit.  Consequently, nodes are more easily distinguished from one another by virtue of their differing neighborhood distributions of labels.

One puzzle remains.  There is no consistent discernible impact of homophily, as measured by the inbreeding coefficient, on vulnerability.  Our procedure for investigating the impact of homophily simply involves swapping labels without disturbing ties.  Therefore, both local and global (unlabeled) topologies remain constant as we decrease the number of cross-group ties to achieve a target value implied by a particular inbreeding coefficient for a given proportional split along the binary attribute.  This procedure disturbs the local labeled topology but because the machine learning attack uses information from that local topology it apparently can adapt to the changes and make equally successful predictions regardless of the value of the inbreeding coefficient.  Perhaps that is why many different factors in attacks on the labeled graphs have some degree of responsibility for success and, no relatively small subset gets the lion's share of the credit.

% In fact, more nodes are distinguishable from the identities with respect to attributes. 
% We confirm this phenomenon from our results even with a binary valued attribute attached to nodes.
% Further, we show this loss of anonymity is amplified when incorporating the attribute information of a $q$ hop neighborhood (i.e., NAD) available to an attacker as background information.

% \clearpage

\bibliographystyle{plain}
\bibliography{bib-v1}

\begin{thebibliography}{10}

\bibitem{adamic2005political}
Lada~A Adamic and Natalie Glance.
\newblock The political blogosphere and the 2004 us election: divided they
  blog.
\newblock In {\em Proceedings of the 3rd international workshop on Link
  discovery}, pages 36--43. ACM, 2005.

\bibitem{aggarwal2011hardness}
Charu~C Aggarwal, Yao Li, and S~Yu Philip.
\newblock On the hardness of graph anonymization.
\newblock In {\em Data Mining (ICDM), 2011 IEEE 11th International Conference
  on}, pages 1002--1007. IEEE, 2011.

\bibitem{backstrom2007wherefore}
Lars Backstrom, Cynthia Dwork, and Jon Kleinberg.
\newblock Wherefore art thou r3579x?: anonymized social networks, hidden
  patterns, and structural steganography.
\newblock In {\em Proceedings of the 16th international conference on World
  Wide Web}, pages 181--190. ACM, 2007.

\bibitem{blackburn2014}
Kourtellis N. Skvoretz J. Ripeanu~M. Blackburn, J. and A.~Iamnitchi.
\newblock Cheating in online games: A social network perspective.
\newblock {\em ACM Transactions on Internet Technology}, 13(3):9:1--9:25, 2014.

\bibitem{smote}
Nitesh~V. Chawla, Kevin~W. Bowyer, Lawrence~O. Hall, and W.~Philip Kegelmeyer.
\newblock Smote: Synthetic minority over-sampling technique.
\newblock {\em J. Artif. Int. Res.}, 16(1):321--357, June 2002.

\bibitem{gong2014joint}
Neil~Zhenqiang Gong, Ameet Talwalkar, Lester Mackey, Ling Huang, Eui
  Chul~Richard Shin, Emil Stefanov, Elaine~Runting Shi, and Dawn Song.
\newblock Joint link prediction and attribute inference using a
  social-attribute network.
\newblock {\em ACM Transactions on Intelligent Systems and Technology (TIST)},
  5(2):27, 2014.

\bibitem{griffith2005messin}
Virgil Griffith and Markus Jakobsson.
\newblock Messin'with texas deriving mother's maiden names using public
  records.
\newblock In {\em Applied Cryptography and Network Security}, pages 91--103.
  Springer, 2005.

\bibitem{gulyas2016efficient}
G{\'a}bor~Gy{\"o}rgy Guly{\'a}s, Benedek Simon, and S{\'a}ndor Imre.
\newblock An efficient and robust social network de-anonymization attack.
\newblock In {\em Proceedings of the 2016 ACM on Workshop on Privacy in the
  Electronic Society}, pages 1--11. ACM, 2016.

\bibitem{haas2016data}
Peter~J Haas.
\newblock Data-stream sampling: basic techniques and results.
\newblock In {\em Data Stream Management}, pages 13--44. Springer, 2016.

\bibitem{handcock2014statnet}
M~Handcock, David~R Hunter, Carter~T Butts, S~Goodreau, P~Krivitsky, Skye
  Bender-deMoll, and Martina Morris.
\newblock statnet: Software tools for the statistical analysis of network data.
\newblock {\em The Statnet Project (http://www. statnet. org). R package
  version}, 2014.

\bibitem{henderson2011s}
Keith Henderson, Brian Gallagher, Lei Li, Leman Akoglu, Tina Eliassi-Rad,
  Hanghang Tong, and Christos Faloutsos.
\newblock It's who you know: graph mining using recursive structural features.
\newblock In {\em Proceedings of the 17th ACM SIGKDD international conference
  on Knowledge discovery and data mining}, pages 663--671. ACM, 2011.

\bibitem{holland1981exponential}
Paul~W Holland and Samuel Leinhardt.
\newblock An exponential family of probability distributions for directed
  graphs.
\newblock {\em Journal of the american Statistical association},
  76(373):33--50, 1981.

\bibitem{hunter2008ergm}
David~R Hunter, Mark~S Handcock, Carter~T Butts, Steven~M Goodreau, and Martina
  Morris.
\newblock ergm: A package to fit, simulate and diagnose exponential-family
  models for networks.
\newblock {\em Journal of statistical software}, 24(3):nihpa54860, 2008.

\bibitem{JiMittal2016De-SAG}
S.~Ji, T.~Wang, J.~Chen, W.~Li, P.~Mittal, and R.~Beyah.
\newblock De-sag: On the de-anonymization of structure-attribute graph data.
\newblock {\em IEEE Transactions on Dependable and Secure Computing},
  PP(99):1--1, 2017.

\bibitem{ji2015your}
Shouling Ji, Weiqing Li, Neil~Zhenqiang Gong, Prateek Mittal, and Raheem~A
  Beyah.
\newblock On your social network de-anonymizablity: Quantification and large
  scale evaluation with seed knowledge.
\newblock In {\em NDSS}, 2015.

\bibitem{ji2014structural}
Shouling Ji, Weiqing Li, Mudhakar Srivatsa, and Raheem Beyah.
\newblock Structural data de-anonymization: Quantification, practice, and
  implications.
\newblock In {\em Proceedings of the 2014 ACM SIGSAC Conference on Computer and
  Communications Security}, pages 1040--1053. ACM, 2014.

\bibitem{ji2016structuralsf}
Shouling Ji, Weiqing Li, Mudhakar Srivatsa, and Raheem Beyah.
\newblock Structural data de-anonymization: Theory and practice.
\newblock {\em IEEE/ACM Transactions on Networking}, 24(6):3523--3536, 2016.

\bibitem{ji2014structure}
Shouling Ji, Weiqing Li, Mudhakar Srivatsa, Jing~Selena He, and Raheem Beyah.
\newblock Structure based data de-anonymization of social networks and mobility
  traces.
\newblock In {\em International Conference on Information Security}, pages
  237--254. Springer, 2014.

\bibitem{ji2016general}
Shouling Ji, Weiqing Li, Mudhakar Srivatsa, Jing~Selena He, and Raheem Beyah.
\newblock General graph data de-anonymization: From mobility traces to social
  networks.
\newblock {\em ACM Transactions on Information and System Security (TISSEC)},
  18(4):12, 2016.

\bibitem{ji2016relative}
Shouling Ji, Weiqing Li, Shukun Yang, Prateek Mittal, and Raheem Beyah.
\newblock On the relative de-anonymizability of graph data: Quantification and
  evaluation.
\newblock In {\em Computer Communications, IEEE INFOCOM 2016-The 35th Annual
  IEEE International Conference on}, pages 1--9. IEEE, 2016.

\bibitem{ji2016survey}
Shouling Ji, Prateek Mittal, and Raheem Beyah.
\newblock Graph data anonymization, de-anonymization attacks, and
  de-anonymizability quantification: A survey.
\newblock {\em IEEE Communications Surveys \& Tutorials}, 2016.

\bibitem{korula2014efficient}
Nitish Korula and Silvio Lattanzi.
\newblock An efficient reconciliation algorithm for social networks.
\newblock {\em Proceedings of the VLDB Endowment}, 7(5):377--388, 2014.

\bibitem{NetflixScandal}
Robert Lemos.
\newblock Researchers reverse netflix anonymization.
\newblock \url{http://www.securityfocus.com/news/11497}, 2007.
\newblock Accessed: 2017-08-11.

\bibitem{leskovec2007dynamics}
Jure Leskovec, Lada~A Adamic, and Bernardo~A Huberman.
\newblock The dynamics of viral marketing.
\newblock {\em ACM Transactions on the Web (TWEB)}, 1(1):5, 2007.

\bibitem{liu2016linkmirage}
Changchang Liu and Prateek Mittal.
\newblock Linkmirage: Enabling privacy-preserving analytics on social
  relationships.
\newblock In {\em NDSS}, 2016.

\bibitem{liu2008towards}
Kun Liu and Evimaria Terzi.
\newblock Towards identity anonymization on graphs.
\newblock In {\em Proceedings of the 2008 ACM SIGMOD international conference
  on Management of data}, pages 93--106. ACM, 2008.

\bibitem{mcdowell2013labels}
Luke~K McDowell and David~W Aha.
\newblock Labels or attributes?: rethinking the neighbors for collective
  classification in sparsely-labeled networks.
\newblock In {\em Proceedings of the 22nd ACM international conference on
  Information \& Knowledge Management}, pages 847--852. ACM, 2013.

\bibitem{mcpherson2001}
Smith-Lovin~L. McPherson, M. and J.~Cook.
\newblock Birds of a feather: Homophily in social networks.
\newblock {\em Annual Review of Sociology}, 27:415--444, 2001.

\bibitem{morris2008specification}
Martina Morris, Mark~S Handcock, and David~R Hunter.
\newblock Specification of exponential-family random graph models: terms and
  computational aspects.
\newblock {\em Journal of statistical software}, 24(4):1548, 2008.

\bibitem{narayanan2011link}
Arvind Narayanan, Elaine Shi, and Benjamin~IP Rubinstein.
\newblock Link prediction by de-anonymization: How we won the kaggle social
  network challenge.
\newblock In {\em Neural Networks (IJCNN), The 2011 International Joint
  Conference on}, pages 1825--1834. IEEE, 2011.

\bibitem{narayanan2009anonymizing}
Arvind Narayanan and Vitaly Shmatikov.
\newblock De-anonymizing social networks.
\newblock In {\em Security and Privacy, 2009 30th IEEE Symposium on}, pages
  173--187. IEEE, 2009.

\bibitem{nilizadeh2014community}
Shirin Nilizadeh, Apu Kapadia, and Yong-Yeol Ahn.
\newblock Community-enhanced de-anonymization of online social networks.
\newblock In {\em Proceedings of the 2014 acm sigsac conference on computer and
  communications security}, pages 537--548. ACM, 2014.

\bibitem{pedarsani2013bayesian}
Pedram Pedarsani, Daniel~R Figueiredo, and Matthias Grossglauser.
\newblock A bayesian method for matching two similar graphs without seeds.
\newblock In {\em Communication, Control, and Computing (Allerton), 2013 51st
  Annual Allerton Conference on}, pages 1598--1607. IEEE, 2013.

\bibitem{pedarsani2011privacy}
Pedram Pedarsani and Matthias Grossglauser.
\newblock On the privacy of anonymized networks.
\newblock In {\em Proceedings of the 17th ACM SIGKDD international conference
  on Knowledge discovery and data mining}, pages 1235--1243. ACM, 2011.

\bibitem{qian2016anonymizing}
Jianwei Qian, Xiang-Yang Li, Chunhong Zhang, and Linlin Chen.
\newblock De-anonymizing social networks and inferring private attributes using
  knowledge graphs.
\newblock In {\em Computer Communications, IEEE INFOCOM 2016-The 35th Annual
  IEEE International Conference on}, pages 1--9. IEEE, 2016.

\bibitem{sala2011sharing}
Alessandra Sala, Xiaohan Zhao, Christo Wilson, Haitao Zheng, and Ben~Y Zhao.
\newblock Sharing graphs using differentially private graph models.
\newblock In {\em Proceedings of the 2011 ACM SIGCOMM conference on Internet
  measurement conference}, pages 81--98. ACM, 2011.

\bibitem{sendina2016assortativity}
I.~Sendi{\~n}a-Nadal, M.~M. Danziger, Z.~Wang, S.~Havlin, and S.~Boccaletti.
\newblock Assortativity and leadership emerge from anti-preferential attachment
  in heterogeneous networks.
\newblock {\em Scientific Reports}, 6:21297 EP --, 02 2016.

\bibitem{sharad2016learning}
Kumar Sharad.
\newblock {\em Learning to de-anonymize social networks}.
\newblock PhD thesis, Computer Laboratory, University of Cambridge, 2016.

\bibitem{Sharad2016benchmark}
Kumar Sharad.
\newblock True friends let you down: Benchmarking social graph anonymization
  schemes.
\newblock In {\em Proceedings of the 2016 ACM Workshop on Artificial
  Intelligence and Security}, AISec '16, pages 93--104, New York, NY, USA,
  2016. ACM.

\bibitem{sharad2013anonymizing}
Kumar Sharad and George Danezis.
\newblock De-anonymizing d4d datasets.
\newblock In {\em Workshop on Hot Topics in Privacy Enhancing Technologies,
  Bloomington, Indiana, USA}, page~10, 2013.

\bibitem{sharad2014automated}
Kumar Sharad and George Danezis.
\newblock An automated social graph de-anonymization technique.
\newblock In {\em Proceedings of the 13th Workshop on Privacy in the Electronic
  Society}, pages 47--58. ACM, 2014.

\bibitem{Skvoretz2013}
John Skvoretz.
\newblock Diversity, integration, and social ties: Attraction versus repulsion
  as drivers of intra- and intergroup relations.
\newblock {\em American Journal of Sociology}, 119:486--517, 2013.

\bibitem{srivatsa2012deanonymizing}
Mudhakar Srivatsa and Mike Hicks.
\newblock Deanonymizing mobility traces: Using social network as a
  side-channel.
\newblock In {\em Proceedings of the 2012 ACM conference on Computer and
  communications security}, pages 628--637. ACM, 2012.

\bibitem{takac2012data}
Lubos Takac and Michal Zabovsky.
\newblock Data analysis in public social networks.
\newblock In {\em International Scientific Conference and International
  Workshop Present Day Trends of Innovations}, number~6 in 1, 2012.

\bibitem{team2014r}
R~Core Team.
\newblock R: A language and environment for statistical computing. vienna,
  austria: R foundation for statistical computing; 2014, 2014.

\bibitem{traud2012social}
Amanda~L Traud, Peter~J Mucha, and Mason~A Porter.
\newblock Social structure of facebook networks.
\newblock {\em Physica A: Statistical Mechanics and its Applications},
  391(16):4165--4180, 2012.

\bibitem{wasserman1996logit}
Stanley Wasserman and Philippa Pattison.
\newblock Logit models and logistic regressions for social networks: I. an
  introduction to markov graphs andp.
\newblock {\em Psychometrika}, 61(3):401--425, 1996.

\bibitem{yartseva2013performance}
Lyudmila Yartseva and Matthias Grossglauser.
\newblock On the performance of percolation graph matching.
\newblock In {\em Proceedings of the first ACM conference on Online social
  networks}, pages 119--130. ACM, 2013.

\end{thebibliography}

% \appendix
% \input{sections/appendix}

\end{document}